\documentclass[journal=jacsat,manuscript=article]{achemso}
\setkeys{acs}{articletitle=true}
\usepackage{setspace}
\usepackage[version=3]{mhchem} 
\usepackage[T1]{fontenc}       
\SectionNumbersOn 
\usepackage{epsfig} 
\usepackage{graphicx} 
\usepackage{amssymb, amstext, amsmath} 
\usepackage{bbold} 
\usepackage{hyperref} 
\usepackage{color}
\usepackage{verbatim}
\usepackage{mathrsfs}
\usepackage[normalem]{ulem} 
\usepackage{esvect}

\title{Deep Learning for Optoelectronic Properties of Organic Semiconductors}
\author{Chengqiang Lu}
\affiliation{Anhui Province Key Lab of Big Data Analysis and Application, University of Science and Technology of China, Hefei, Anhui 230026, China}
\author{Qi Liu}
\email{qiliuql@ustc.edu.cn}
\affiliation{Anhui Province Key Lab of Big Data Analysis and Application, University of Science and Technology of China, Hefei, Anhui 230026, China}
\author{Qiming Sun}
\affiliation{Tencent America, Palo Alto, CA 94306 , United States}
\author{Chang-Yu Hsieh}
\affiliation{Tencent, Shenzhen, Guangdong 518057, China, }
\author{Shengyu Zhang}
\affiliation{Tencent, Shenzhen, Guangdong 518057, China, }
\author{Liang Shi}
\email{lshi4@ucmerced.edu}
\affiliation{Chemistry and Chemical Biology, University of California, Merced, California 95343, United States}
\author{Chee-Kong Lee}
\email{cheekonglee@tencent.com}
\affiliation{Tencent America, Palo Alto, CA 94306 , United States}

\begin{document}
\setstretch{1.25}
\maketitle
\clearpage
\begin{abstract}
Atomistic modeling of energetic disorder in organic semiconductors (OSCs) and its effects on the optoelectronic properties of OSCs requires a large number of excited-state electronic-structure calculations, a computationally daunting task for many OSC applications. In this work, we advocate the use of deep learning to address this challenge and demonstrate that state-of-the-art deep neural networks (DNNs) are capable of predicting the electronic properties of OSCs at an accuracy comparable with the quantum chemistry methods used for generating training data. We extensively investigate the performances of four recent DNNs (deep tensor neural network, SchNet, message passing neural network, and multilevel graph convolutional neural network) in predicting various electronic properties of an important class of OSCs, i.e., oligothiophenes (OTs), including their HOMO and LUMO energies, excited-state energies and associated transition dipole moments. We find that SchNet shows the best performance for OTs of different sizes (from bithiophene to sexithiophene), achieving average prediction errors in the range of 20-80meV compared to the results from (time-dependent) density functional theory. We show that SchNet also consistently outperforms shallow feed-forward neural networks, especially in difficult cases with large molecules or limited training data. We further show that SchNet could predict the transition dipole moment accurately, a task previously known to be difficult for feed-forward neural networks, and we ascribe the relatively large errors in transition dipole prediction seen for some OT configurations to the charge-transfer character of their excited states. Finally, we demonstrate the effectiveness of SchNet by modeling the UV-Vis absorption spectra of OTs in dichloromethane and a good agreement is observed between the calculated and experimental spectra. Our results show the great promise of DNNs in depicting the rugged energy landscapes encountered in OSCs, serving as the first step in the atomistic modeling of optoelectronic processes in OSCs relevant to device performances.
\end{abstract}

\clearpage
\section{Introduction}
Organic semiconductors (OSCs) are a class of conjugated molecules or polymers with promising applications in various optoelectronic devices~\cite{Ostroverkhova2016}, such as solar cells~\cite{Hains2010,Myers2012,Lu2015,Hedley2017}, light-emitting diodes~\cite{Minaev2014,Xu2016}, field-effect transistors~\cite{Sirringhaus2014}, and photo-detectors~\cite{Baeg2013}, due to many of their desirable properties including low weight, flexibility, and easy processability. Their optoelectronic activities rely on the efficient creation, annihilation, or transport of charges (i.e., electrons or holes), or bound electron-hole pairs, namely excitons. Due to the weak van der Waals interactions between organic molecules, substantial configurational disorder is often present in OSCs, which leads to rough energy landscapes for charges and excitons, thereby greatly limiting the efficiencies of organic optoelectronic devices~\cite{Ostroverkhova2016,Moule2008,McMahon2010}. To improve the performances of OSCs, it is essential to understand the energetic disorder in OSCs and their effects on electronic properties, such as charge or exciton transport. Molecular simulations provide a unique avenue to reveal the microscopic molecular disorder and the resulting energetic disorder, usually from molecular dynamics simulations and quantum chemical calculations~\cite{Bredas2009,Troisi2011,Zhugayevych2014}. 

As a critical step in modeling the electronic properties of OSCs, accurate predictions of molecular electronic properties as a function of molecular configuration are essential, and many quantum chemical methods are capable of providing such predictions. For example, time-dependent density functional theory (TDDFT) with properly chosen functionals can often predict the lowest-lying excited-state energy within an error of tenths of eV for most OSCs.\cite{Salzner2011,Jacquemin2015,Faber2014,Kummel2017,Korzdorfer2014} However, the length scales of organic optoelectronic devices (typically tens or hundreds of nanometers) and the time scales of electronic processes in these devices (typically nanoseconds or beyond) require repetitive quantum chemical calculations for a large number of OSC molecules, and most methods, e.g., TDDFT, are still too expensive for these large-scale high-throughput calculations. Instead of performing molecular simulations, many modelers of OSCs resort to phenomenological models~\cite{Feron2012a,Feron2012,Li2013,Tapping2015} with parameters inferred from experiments or a limited number of quantum calculations, e.g., lattice models with Gaussian energetic disorder~\cite{Bassler1993, Shi2017, Lee2016}. These methods have greatly advanced our understanding of electronic processes in OSCs, but have limited predictive power due to the lack of molecular details and often chemical specificity. In this work, we aim at exploring the use of machine learning (ML) in predicting the electronic properties of OSCs with an accuracy comparable to DFT but at a much lower computational cost. 

ML has already found various applications in computational chemistry, such as electronic structure methods~\cite{Ramakrishnan2015, Welborn2018, Zaspel2018, Cheng2019, Brockherde2017, Bogojeski2018, Grisafi2018, Fabrizio2019, Ryczko2019}, force field development~\cite{Huan2017, Smith2017, Botu2016, Chmiela2017, Chmiela2018, Wang2019, Jinnouchi2019}, spectroscopy\cite{Ramakrishnan2015b,Ye2019, Gastegger2017, Ghosh2019} and virtual screening~\cite{Jorissen2005, Gomez-Bombarelli2016, Gomez-Bombarelli2018}. Due to the recent availability of standard datasets and increasing interest in extending ML methods to graph data~\cite{2019Wu}, there have been much effort within both the chemistry and ML communities to develop deep neural networks (DNNs) that treat molecules as computational graphs~\cite{Duvenaud2015, Schutt2017, Gilmer2017, Schutt2018, Wu2018, Unke2019, Schutt2019, Lu2019a, Chen2019}. In contrast with traditional ML methods where hand-crafted molecular descriptors, such as Coulomb matrix~\cite{Rupp2012,Hansen2013}, are required as input, deep learning approaches are capable of extracting the optimal representation of a molecule solely from atom types and Cartesian coordinates. State-of-the-art DNNs have shown impressive performance on benchmark datasets such as the QM9 dataset which contains 133,885 small organic molecules consisting of up to 9 non-hydrogen atoms (i.e., carbon, nitrogen, oxygen and fluorine)~\cite{Ramakrishnan2014}, often achieving chemical accuracy on a variety of ground state properties~\cite{Wu2018, Chen2019}. Despite these successes, two critical components are still absent before these advanced DNNs could be widely adopted for OSCs. First, these DNNs have not been systematically tested on molecules larger than those found in the QM9 dataset, thus it remains unclear if such high accuracy is transferable to larger molecules, such as many OSCs.  Second, most benchmarking effort focuses on predicting the ground-state properties such as the highest-occupied-molecular-orbital (HOMO) energy, lowest-unoccupied-molecular-orbital (LUMO) energy and dipole moment~\cite{Wu2018,Chen2019}, and there have been limited studies on applying these advanced DNNs to excited-state properties such as excitation energies and the associated transition dipoles, which are crucial in determining the device performance of OSCs~\cite{St.John2019}. Excited-state properties could be very different from ground-state properties, e.g., the electron correlation in excited states could be much stronger than that in ground state~\cite{Fulde1995}, thus one cannot simply expect that the high accuracy of DNNs in ground-state predictions is directly transferable to excited-state properties.

In this paper, we address these issues by using oligothiophenes (OTs) as model systems and performing a comprehensive investigation of the performances of four state-of-the-art DNN models on ground- and excited-state electronic properties that are directly related to the optoelectronic applications of OTs. OTs are one of the most studied oligomeric OSCs, and have important applications in thin film devices, such as field-effect transistors~\cite{Fichou1999,Perepichka2009}. More importantly, OTs serve as excellent finite model systems for thiophene-based polymers due to their well-defined chemical structures and higher processability compared to the corresponding polymers~\cite{Fichou1999}. It has been long recognized that the electronic properties of polythiophenes may be inferred or extrapolated from those of OTs~\cite{Fichou1992,DeMelo1999,Fabiano2005}. The specific OTs considered here are $\alpha$-linked OTs, denoted as $n$T, where $n$=2-6 stands for the number of thiophene rings in the oligomer. The specific electronic properties of OTs considered here include the HOMO and LUMO energies, the input properties for charge-transport modeling, the HOMO-LUMO energy gap and electronic excited-state energy, two popular descriptors for exciton energy.

After extensively benchmarking four DNN models against quantum chemical results of these properties,  we find that all the DNN models retain their impressive performance even as molecular size increases, achieving mean average errors (MAEs) significantly below 0.1eV for HOMO energy, LUMO energy, HOMO-LUMO gap and excited-state energies. Among the DNNs tested, SchNet is found to be the most accurate model for all tested properties. It is observed that while the accuracy of SchNet is only slightly higher than that of a shallow neural network for small molecules, SchNet significantly outperforms shallow neural networks in predicting the electronic properties of larger molecules. As opposed to speech and image recognition where DNNs benefit most from data abundance, the performance advantage of SchNet over shallow neural networks is even more drastic when training data are limited. These findings suggest that DNNs are especially powerful for difficult cases such as those involve large molecules and limited training data. Besides energetic properties, we also use SchNet to predict the magnitude of transition dipole moment, which has been previously shown to be difficult for shallow feed-forward neural network~\cite{Ye2019}, and obtain satisfactory accuracy. Moreover, we show that molecular configurations with large errors in predicted transition dipole moment are often associated with large electron-hole separations, indicating that one should exercise caution when applying ML methods for systems with strong charge-transfer character. Finally, we demonstrate the validity and effectiveness of SchNet in predicting the excited-state properties of OTs by modeling the UV-Vis absorption spectra of OTs in dichloromethane, and a good agreement between calculated and experimental spectra is achieved.  

The organization of the rest of this paper is as follows. In Section \ref{sec:data_generation}, the general workflow and the simulation details for data generation are provided; In Section \ref{sec:model_selection}, four DNN models are briefly reviewed, and their performances in predicting electronic properties of OTs are compared; In Section \ref{sec:model_validation}, the best-performing DNN model, SchNet, is further scrutinized, and physical insights are provided to its prediction errors. In Section \ref{sec:spec}, absorption spectra of OTs in dichloromethane are computed using SchNet, and direct comparison is made between experiment and calculation. Finally, in Section \ref{sec:conclusion}, we present our concluding remarks.

\begin{figure}[h!tbp]
  \centering
    \includegraphics[width=1.0\textwidth]{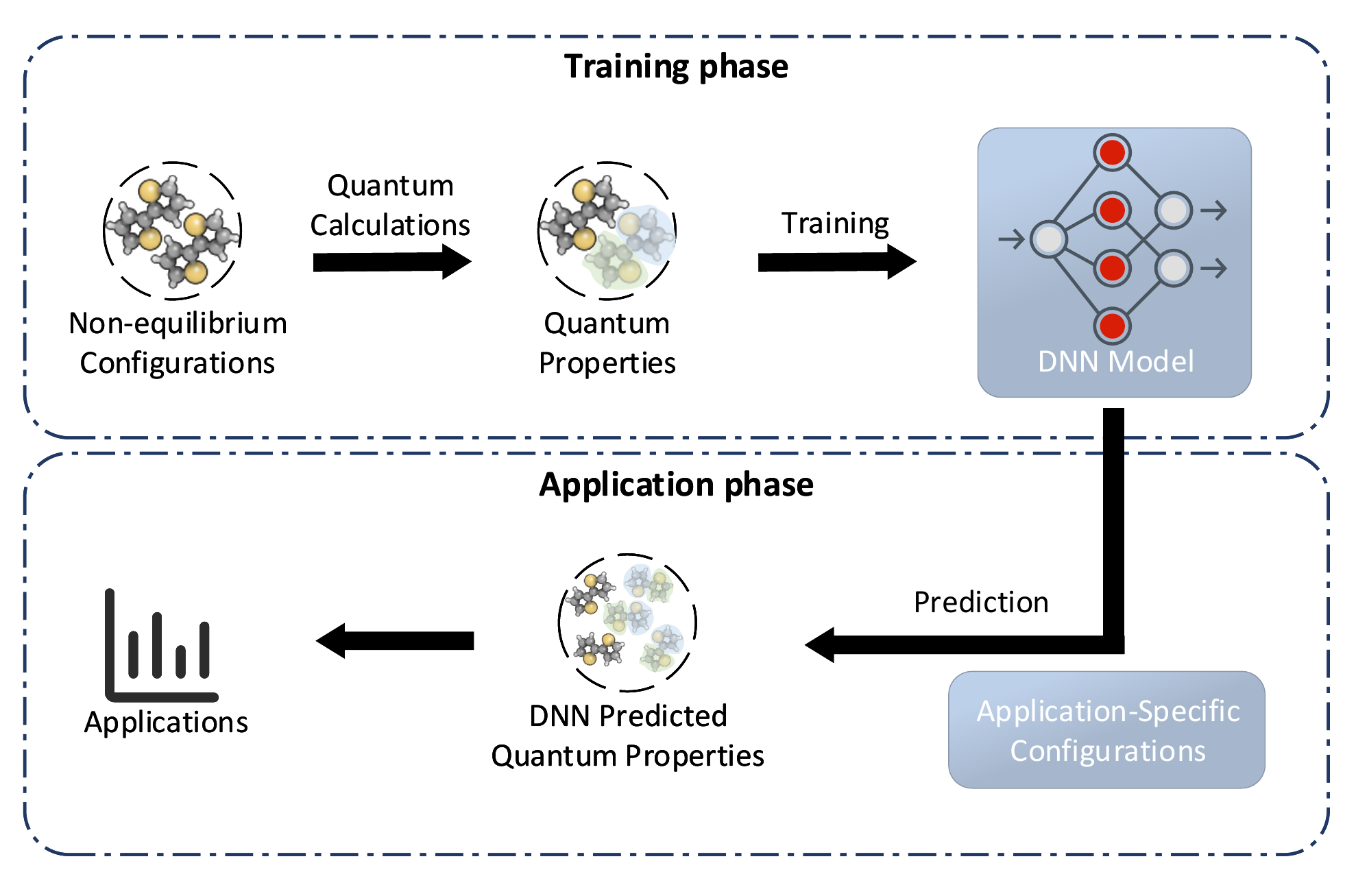}
     \caption{
     A schematic of the ML protocol used in this work. 
     \textbf{Training phase:} 
     High-temperature gas-phase MD simulations are used to generate non-equilibrium molecular configurations. Quantum chemistry calculations are then performed to obtain the quantum properties of these configurations, which are used to train DNN models. 
     \textbf{Application phase:} Configurations of OSCs under experimental conditions are used as input for the trained DNN models, and the predicted electronic properties can then be used for specific application, e.g., computing absorption spectra.}
     \label{fig:workflow}
\end{figure}

\section{Data Generation}\label{sec:data_generation}

Fig. \ref{fig:workflow} illustrates the general workflow of our ML protocol for modeling OSCs, which consists of the training phase and application phase. In the training phase, a large number of non-equilibrium molecular configurations are generated from molecular simulations, e.g., high-temperature gas-phase molecular dynamics (MD) simulation. Quantum chemical calculations are then performed on these configurations to generate datasets of the quantum properties of interest, e.g., excite-state energies, and DNN models are trained to relate configurations and quantum properties. In the application phases, application-specific configurations, e.g., configurations of OTs in solutions, are fed into the DNN models to obtain predicted quantum properties at a much lower computational cost compared to quantum chemical calculations. Finally, the DNN-predicted quantum properties are used to compute observables for specific applications, e.g., absorption spectra of OTs in solutions.  

In this work, the non-equilibrium configurations of OTs were harvested from classical MD simulations using the OPLS/2005 force field~\cite{Banks2005}. It is known that the excited state properties of OTs are greatly affected by the torsional motions in OTs~\cite{Fave1992,Pan2002,Westenhoff2006a,Darling2008}, and OPLS/2005 was chosen here in that the torsional potential energy surface of 2T from OPLS/2005 has been shown to reasonably reproduce that from localized second-order Moller-Plesset perturbation theory (LMP2)~\cite{Dubay2012}. For the simulations of isolated OTs, a single OT molecule was placed in a simulation box much larger than the size of OT. All the MD simulations were then performed with the Desmond package 3.6~\cite{Bowers2006} in the NVT ensemble at 1000K to ensure adequate sampling of OT configurations, particularly the high-energy ones. Here we aim to develop transferable DNNs that, once trained, work for OTs in different molecular environments including solution phases, crystalline and amorphous solid phases at varying temperatures. Nos\'{e}-Hoover thermostat~\cite{nose84b,hoover85} with a coupling constant of 2.0ps was employed to maintain the temperature, and the electrostatic interactions were computed using the particle-mesh Ewald method~\cite{darden93,essmann95}. The simulation time step was 1fs, and configurations were saved every 100fs over a 10ns simulation for each OT.

To train DNN models for electronic properties of OTs, we compiled a dataset for each OT by employing quantum chemistry on 100,000 OT configurations generated from MD simulation. Density functional theory (DFT) with CAM-B3LYP functional and 6-31+G(d) basis set was employed to compute HOMO and LUMO energies, and time-dependent DFT (TDDFT) calculations were performed within the Tamm-Dancoff approximation to obtain the singlet excited-state energies and associated transition dipoles. The range-separated functional, CAM-B3LYP, was chosen to reduce the self-interaction error~\cite{Salzner2011}, and the inclusion of the diffuse function in the basis set has been shown necessary for OTs~\cite{Sun2014}. The accuracy of CAM-B3LYP/6-31+G(d) in predicting the lowest singlet excited-state energies of OTs is validated against the results from a higher-level theory (CC2), as discussed in Section II of Supplementary Information (SI). All the quantum chemical calculations were performed using the PySCF program~\cite{Sun2018,Sun2015}. The datasets of the electronic properties of OTs are then used to train and evaluate four DNN models in the next subsection.

\section{Model Selection}\label{sec:model_selection}
Unlike conventional ML methods, DNNs are capable of automatically extracting optimal representations from molecular configurations without resorting to the more traditional approach of manually designing descriptors such as Coulomb matrices~\cite{Rupp2012, Hansen2013}, bags of bonds~\cite{Hansen2015}, smooth overlap of atomic positions~\cite{Bartok2013} or generalized symmetry functions~\cite{Behler2011}. In this section, we consider four state-of-the-art DNNs for molecules and evaluate their performances in predicting the ground- and excited-state properties of OTs.

\noindent \textbf{Deep Tensor Neural Network (DTNN)}\cite{Schutt2017}: In DTNN, molecules are encoded by a vector of nuclear charges and an inter-atomic distance matrix, such that the input is rotationally and translationally invariant. Inspired by quantum many-body physics, DTNN consists of many interaction blocks in order to mimic the interactions of atoms. Additionally, DTNN provides a systematic approach to partitioning extensive molecular properties into atomic contributions by predicting the individual contribution of each atom in a molecule.

\noindent \textbf{SchNet}\cite{Schutt2018, Schuett2017a}: SchNet shares many structural similarities with DTNN, namely the translationally and rotationally invariant molecular representation and the concept of interaction layers. The developers of SchNet proposed to use continuous convolution filters that are able to handle unevenly spaced data, particularly atoms at arbitrary positions, as opposed to traditional convolution neural networks where the input data consists of discrete image pixels. By adding the gradients of energy into the loss function, SchNet is also able to predict energy conserving force field with high precision~\cite{Schutt2018, Schuett2017a}.

\noindent \textbf{Message Passing Neural Network (MPNN)}\cite{Gilmer2017}: Treating a molecule as a computational graph, MPNN uses bond (as edge) and atom (as node) type features in addition to the interatomic distances. MPNN consists of two phases: a message passing phase and a readout phase. Message passing phase is used to extract information of the molecular graph, whereas the readout phase is responsible for mapping the graph to its properties. There are multiple variants of MPNN, and in this paper we only consider the best performing variant which uses an edge network for message passing and a set2set function for readout. MPNN has recently been used for high-throughput screening of polymeric organic photovoltaic applications materials~\cite{St.John2019}.

\noindent \textbf{Multilevel Graph Convolutional neural Network (MGCN)}\cite{Lu2019a}: Similar to MPNN, each molecule in MGCN is represented as a computational graph and the molecular representation is passed through a series of interaction layers in the message passing phase. Within the interaction layers in the message passing phase, the inter-atomic interactions are modeled in a hierarchical fashion in order to capture the many-body interactions (two-body, three-body descriptors, etc.). Additionally, the readout function uses the representation of all the interaction layers instead of only the last interaction layer as in MPNN.

All the above models have shown impressive results when tested on datasets consisting of small organic molecules, such as the QM9 dataset which contains the equilibrium geometries and ground-state properties of 133,885 small organic molecules with up to 9 non-hydrogen atoms (C, N, O, F)~\cite{Ramakrishnan2014}. In the QM9 dataset, these DNNs often achieve an average error smaller than 1 kcal/mol for a wide range of ground-state properties. However, to our knowledge, these state-of-the-art DNNs have not been systematically tested on excited states or larger molecules, thus their performance remains unclear for many applications in OSCs. In the following, we provide such a comparison and investigate the performance of these models in predicting the electronic properties of OTs of different sizes.

The 100,000 MD configurations of each OT described in Section~\ref{sec:data_generation} are randomly split into training, validation, and test sets containing 80000, 10000, and 10000 configurations, respectively. The validation set is used to pick the optimal number of epochs for early stopping in the training phase, and we use the same hyperparameters as the original papers unless otherwise stated. The implementation details of the models are described in Section I of the SI. The source codes of our implementations of SchNet, MPNN and MGCN can be found on Github~\cite{alchemy}, whereas DTNN is implemented using the DeepChem library~\cite{Ramsundar2019}. 

\begin{figure}[h!tbp]
  \centering
    \includegraphics[width=0.9\textwidth]{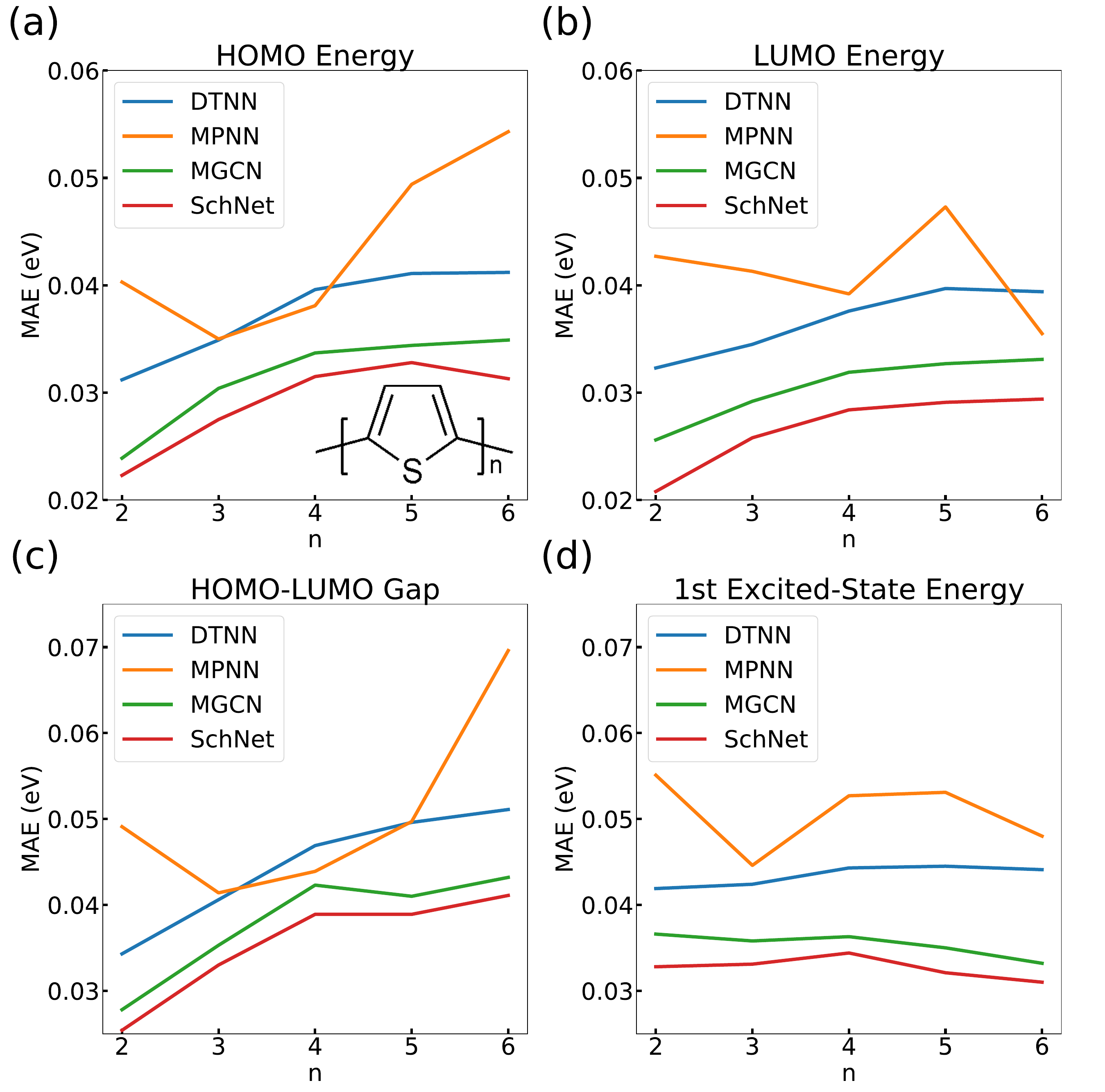}
     \caption{We use oligothiophenes (OTs) of varying lengths to test the performance of four state-of-the-art DNNs in predicting various optoelectronic properties. Plotted above are the test set mean average errors (MAEs) as a function of the length of OTs, and the optoelectronic properties considered here are (a) HOMO energy, (b) LUMO energy, (c) HOMO-LUMO gap and (d) first excited-state energy.}
     \label{fig:models_comparison}
\end{figure}

We evaluate the performances of the selected DNNs on HOMO energy, LUMO energy, HOMO-LUMO gap, and the first excited-state energy of OTs, and we train one model for each property. The resulting mean average errors (MAEs) on the test data are shown in Fig.~\ref{fig:models_comparison}. It is found that in general the MAEs increase slightly with OT length for ground-state properties, but remain approximately constant for excited-state energy prediction. Overall, all DNNs produce satisfactory results with MAEs ranging approximately from 20meV to 70meV (20meV to 50meV after excluding MPNN). Such MAEs are comparatively small given the range of fluctuations in the data, which is in the order of several eV. For example, the ranges of HOMO energy, LUMO energy, HOMO-LUMO gap and excited-state energy of 6T in the data are 1.9eV, 1.6eV, 3.1eV and 2.9eV, respectively. It is also worth noting that these MAEs are also minute compared to the typical errors of DFT methods, generally considered to be in the range of several hundred meV~\cite{Dreuw2005}. The observation that the performances of DNNs do not degrade significantly with system size indicates that these DNNs may be applied to larger molecular systems without significant loss of accuracy.

From Fig.~\ref{fig:models_comparison}, it is clear that SchNet consistently outperforms other DNN models on all properties and system sizes. This suggests that the continuous convolutional filters employed in SchNet are most suited to capturing how minute changes in atomic positions affect the electronic properties of OTs. Despite this, the MAEs of MGCN are only marginally higher than SchNet in all our numerical experiments, indicating that MGCN could also be the DNN of choice for optoelectronic property predictions. From Fig.~\ref{fig:models_comparison}, we found that the MAEs of MPNN exhibit a large variation, consistent with earlier findings on QM9 dataset~\cite{Wu2018}. In the subsequent sections, we will use SchNet for more validation experiments before proceeding to compute absorption spectrum which involves predicting the energies and transition dipole moments of the first few excited states.

\section{Model Validation}\label{sec:model_validation}
\begin{figure}[h!tbp]
  \centering
    \includegraphics[width=0.9\textwidth]{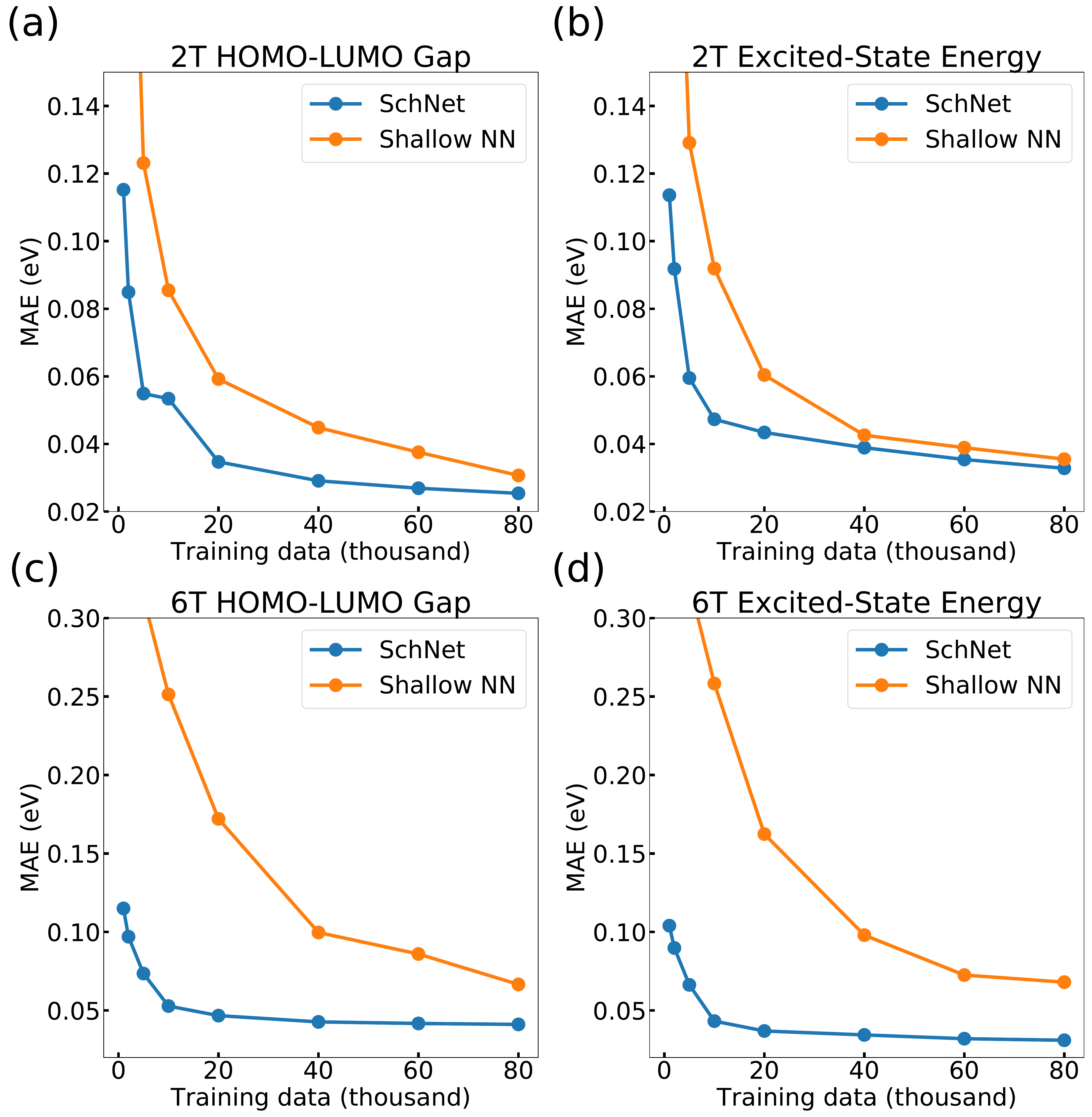}
     \caption{The test set MAEs of SchNet (solid blue lines) in predicting the HOMO-LUMO gap (left panel) and first excited-state energy (right panel) of 2T and 6T as a function of the number of training data. 
     For comparison, the corresponding results using a shallow feed-forward neural network with 2 hidden layers and Coulomb matrix as input are included (solid orange lines). The first 4 points of the figures correspond to training data sizes of 1000, 2000, 5000 and 10000, respectively.}
     \label{fig:training_size}
\end{figure}

In this section, we perform more numerical experiments to further evaluate the performance of SchNet. We first investigate how MAEs depend on the size of the training set in Fig.~\ref{fig:training_size}. For conciseness, we only include the results of 2T and 6T in Fig.~\ref{fig:training_size} since we found that the training-set size dependence of SchNet for 2T-6T to be quantitatively similar. It can be seen that even with a training set of only 5000 data points, the MAEs for HOMO-LUMO gap and first excited-state energy are already significantly lower than 0.1eV. 
Beyond 20,000 training data points, additional training data only leads to modest improvement. Given that the generation of training data could be the bottleneck for many ML applications in chemistry, the results in Fig.~\ref{fig:training_size} could provide a rough guideline for estimating the amount of training data needed given a desired error tolerance. 

We next evaluate the performance of SchNet on other excited-state properties before proceeding to compute the absorption spectrum in Section~\ref{sec:spec}. The MAEs for the first 5 excited-state energies and the associated transition dipole magnitude squared, $|\mu_i|^2$, are listed in Table \ref{tab:t6}. It is observed that the MAEs of excited-state energies generally increase as we go to higher excited states. However, the accuracy in the energy prediction remains satisfactory considering the range of fluctuations in the datasets (several eV) and the typical errors in TDDFT (tenths of eV). For example, the fifth excited-state energy has the largest MAE of 78meV, which is only a $3.3\%$ relative error given the range of fluctuation is 2.4eV. On the other hand, the MAEs for $|\mu_i|^2$ decreases as we go to higher excited states. This is due to the decrease of the average magnitude and range of fluctuation of $|\mu_i|^2$ for higher excited states. For example, using the cutoff of 5\AA\, the MAEs of predicted $|\mu_i|^2$ of the first excited state is about 1.1$D^2$ compared to the fluctuation range of 40.1$D^2$, whereas for the fifth excited state the corresponding MAE is only 0.29$D^2$ compared to the fluctuation range of $7.0D^2$.

\begin{table}[h!tbp]
\begin{tabular}{ccccc}
\hline
 &\multicolumn{2}{c}{cut-off=5\AA} &  \multicolumn{2}{c}{cut-off=25\AA} \\
 \hline
excited state $i$ & energy & $|\mu_i|^2$  & energy & $|\mu_i|^2$  \\
\hline
1             & 0.031 & 1.060 & 0.031 & 0.713                      \\
2             & 0.058 & 1.031 & 0.050 & 0.776                      \\
3             & 0.065 & 0.654 & 0.064 & 0.626                      \\
4             & 0.061 & 0.430 & 0.061 & 0.432                      \\
5             & 0.078 & 0.294 & 0.078 & 0.288     \\
\hline
\end{tabular}
\caption{The test set MAEs of SchNet in predicting the first 5 excited-state energies (in eV) and the associated transition dipole moment squared, $|\mu_i|^2$, of 6T using cutoff radii of 5\AA\; and 25\AA.}
\label{tab:t6}
\end{table}

In order to model the interactions of atoms and restrain the neural network, SchNet introduces a cutoff distance, beyond which the interaction between two atoms decreases rapidly following a Gaussian function. A larger cutoff radius enables DNNs to take into account long-range interactions more effectively albeit at the expense of higher computational cost. For example in our implementation, the training time increases from 30 seconds to 55 seconds per epoch on a single Nvidia V100 GPU as we increase the cutoff radius from 5\AA\ to 25\AA. In Table \ref{tab:t6}, we find that the accuracy of transition dipole prediction could be further improved by using a larger cutoff radius. Increasing the cutoff from 5\AA\ to 25\AA, approximately the length of the entire 6T molecule, the MAEs for the $|\mu_i|^2$ of the first two excited states drop by $33\%$ and $25\%$, respectively, whereas the MAEs for excited-state energies remain largely unaltered. The sensitivity of transition dipole prediction to cutoff radius is a result of the charge-transfer character in some molecular configurations and will be discussed in detail in Section \ref{sec:CT_state}. It is worth noting that the prediction of transition dipole moment has been shown to be difficult for conventional feed-forward neural networks~\cite{Ye2019} unless information from natural population analysis~\cite{Reed1985} (NPA) is used in the molecular representation. However, NPA itself requires ground-state electronic structure calculations, thus this might not be practical for applications involving large molecules or many molecular configurations.

\subsection{Deep versus shallow neural networks}
Training DNNs typically requires expensive hardware (e.g., GPUs) and long training time. For example, the training times of the models in this work with 80,000 6T configurations range from 4 to 20 hours on a single Nvidia V100 GPU, depending on the specific DNN model. Despite the high training cost, we find that DNNs are still preferred over traditional ML methods given their superior performance. To demonstrate the performance advantage of DNNs, we include the results from shallow feed-forward neural networks in Fig.~\ref{fig:training_size} (orange solid lines) for comparison. The feed-forward neural network consists of two hidden layers, each hidden layer contains 200 hidden units and Coulomb matrix is used as the input (the expression of the Coulomb matrix can be found in Ref. \citenum{Hansen2013}). We find that more hidden layers or hidden nodes do not lead to noticeably better results.

Several remarks are worth noting from the comparison between SchNet and feed-forward neural network. First, MAEs from SchNet are always lower than those from feed-forward neural network for both HOMO-LUMO gap and excited state energy, regardless of training set size, confirming that DNNs provide superior performance compared to traditional ML methods where hand-crafted features are required. Second, the performance advantage of SchNet is more prominent for 6T than for 2T, suggesting that DNNs are even more suitable for predicting the electronic properties of large molecules. Third, in cases when training data are limited (e.g., less than 10,000), SchNet still retains good predictive power whereas feed-forward neural networks fail to provide reasonable estimates under such restriction. For example, MAEs from SchNet remain less than $0.12eV$ for all the properties in Fig.~\ref{fig:training_size} even with only 1,000 training data, but the MAEs of feed-forward neural network can be as large as $0.48eV$ for excited-state energies of 2T and 6T.

\subsection{Transition Dipole Moment Prediction and Charge-Transfer Character}\label{sec:CT_state}
\begin{figure}[h!tbp]
  \centering
    \includegraphics[width=0.9\textwidth]{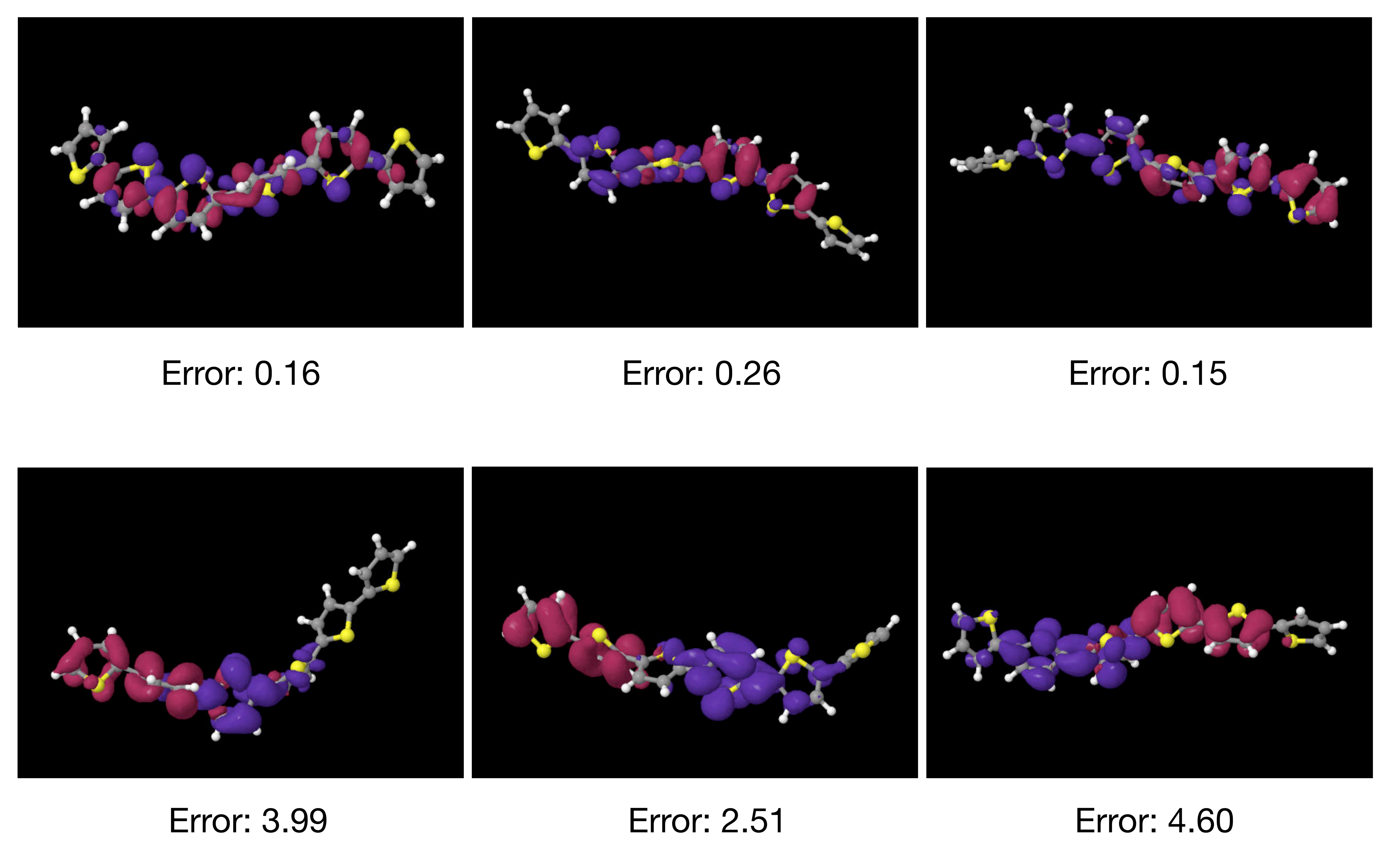}
     \caption{
     Representative 6T molecular configurations with small (top row) and large (bottom row) errors in SchNet predicted $|\mu_1|^2$, in units of $D^2$. 
     Electron density difference between the ground and first excited states is shown as blue (red) 0.001 iso-surface representing electron density gain (loss)}
     \label{fig:density_diff}
\end{figure}

\begin{figure}[h!tbp]
  \centering
    \includegraphics[width=0.7\textwidth]{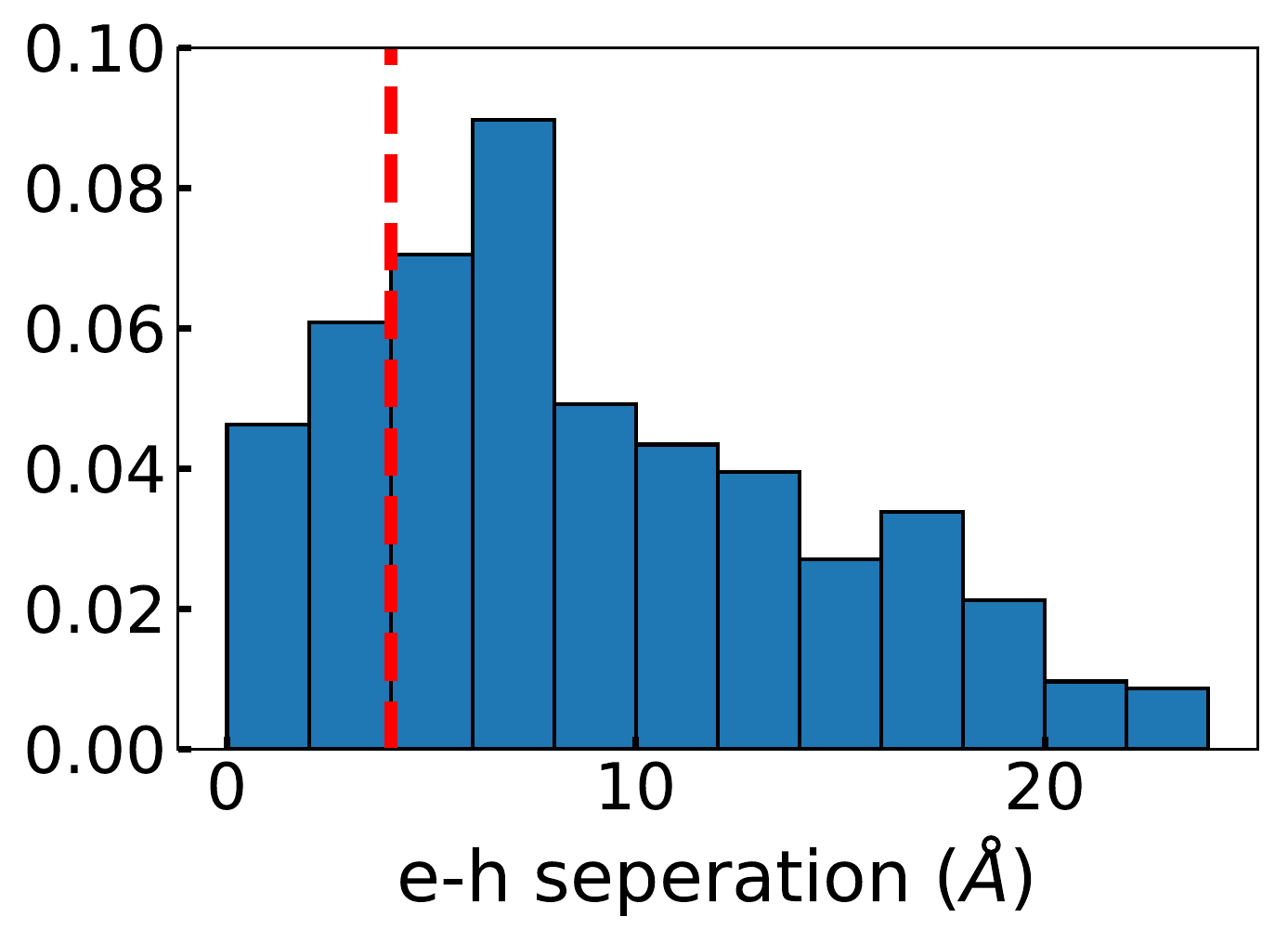}
     \caption{Histogram of electron-hole separation for 6T molecular configurations with errors in SchNet predicted $|\mu_1|^2$ greater than 2.0$D^2$, using a cutoff radius of 25\AA. Among these configurations, $79\%$ of them have electron-hole separation larger than 4\AA\, (approximately the length of one thiophene ring, red dashed line). }
     \label{fig:e-h-seperation}
\end{figure}

Transition dipole moment depends on both the ground and excited-state wavefunctions
\begin{eqnarray}
\mu_i = \langle i | \hat{\mu} | 0 \rangle,
\end{eqnarray}
where the $|0\rangle$ and $|i\rangle$ represent the ground and the $i$th excited states, respectively, $\hat{\mu}$ is the dipole operator, and $\mu_i$ is the transition dipole moment associated with the $i$th excited state. In Table \ref{tab:t6} we show that the MAEs for transition dipole moment predictions are sensitive to the cutoff radius, suggesting that the change in electron density from the ground state to excited state is highly non-local. To visualize the change in electron density between the ground and first excited states, we randomly choose several 6T configurations with either small or large errors in the predicted $|\mu_1|^2$ and their electron density differences are shown in Fig.~\ref{fig:density_diff}. The top row in Fig.~\ref{fig:density_diff} shows three 6T configurations with small errors in predicted $|\mu_1|^2$, whereas the bottom row shows three 6T configurations with large errors. The blue (red) 0.001 iso-surface shows the electron density gain (loss) associated with the first excited state of 6T. It can be seen from Fig.~\ref{fig:density_diff} that configurations with large errors also exhibit strong charge-transfer character, but such charge-transfer character is much less evident in configurations with small transition dipole errors. This observation indicates some correlation between the prediction accuracy of transition dipole moment and charge-transfer character of the excited state. To further examine this correlation, we look at the distribution of average electron-hole separations for 6T configurations with large errors in predicted $|\mu_1|^2$ (larger than 2.0$D^2$), shown in Fig.~\ref{fig:e-h-seperation}. It can be seen that the electron and hole are well separated for most configurations with large prediction errors. For example, $79\%$ of these configurations have electron-hole separations greater than 4\AA, i.e., roughly the length of one thiophene ring. Our analysis here suggests careful assessment must be taken in using ML methods for calculating transition dipole moments for molecules with strong charge transfer character.

\section{Application to Absorption Spectrum}\label{sec:spec}

To demonstrate the capability of SchNet, we compute the UV-Vis absorption spectra for OTs in dichloromethane, possibly the most commonly measured electronic property of OSCs. Despite recent progress~\cite{Segatta2019,Loco2019,Zuehlsdorff2019a,Zuehlsdorff2019,Zuehlsdorff2018}, accurate atomistic modeling of solution-phase absorption spectra remains challenging, and one of the challenges is the large number of configurations responsible for spectral inhomogeneity, whose excited-state properties need to be computed via quantum chemistry methods. The relatively high computational cost of excited-state calculations calls for more efficient methods, such as many semi-empirical methods~\cite{Shi2018}, 
and here we will show that DNNs can also provide predictions at an accuracy comparable to the targeted quantum chemistry method. 

For each OT, one million configurations were generated from the MD simulation of OT in dichloromethane at 300K and 1atm, as detailed in Section III of SI, and they were then fed into the SchNet models to compute the energies and transition dipoles for the two lowest-lying singlet excited states. The absorption spectra of OTs in solutions are expected to be dominated by inhomogeneous broadening, and we computed the absorption coefficient, $\alpha(E)$, in the inhomogeneous limit, given by\cite{mcquarrie76,mukamel95}
\begin{equation}
    \alpha(E) \sim E\sum_{i=1,2} \langle |\mu_i |^2 \delta(E-\Omega_i)\rangle,
    \label{eq:abs}
\end{equation}
where $\delta(x)$ is the Kronecker delta function, and $\Omega_i$ and $\mu_i$ are the transition energy and transition dipole moment associated with the $i$th excited state, respectively. The angular brackets indicate an ensemble average, which is equivalent to a time average over the 10-ns trajectory. Note that in Eq. (\ref{eq:abs}), we have ignored the factor, $1-e^{-E/k_B T}$, where $k_B$ is the Boltzmann factor and $T$ is the temperature,  as for electronic spectroscopy, $E \gg k_B T$. In using Eq. (\ref{eq:abs}), we have also assumed the adiabaticity between the two excited states, a good approximation for OTs in solutions~\cite{Becker1996,Egelhaaf1998}. Another caveat is that the solvent effects on the excited-state energies and transition dipoles are neglected, and this approximation is acceptable given that the solvatochromism of OTs is relatively small (see Section III in SI for more discussions)~\cite{Taliani,Grebner1995,Lap1997,Zhao1988,Sease1947,Salzner2007,Aleman2011}.

\begin{figure}[h!tbp]
  \centering
    \includegraphics[width=0.5\textwidth]{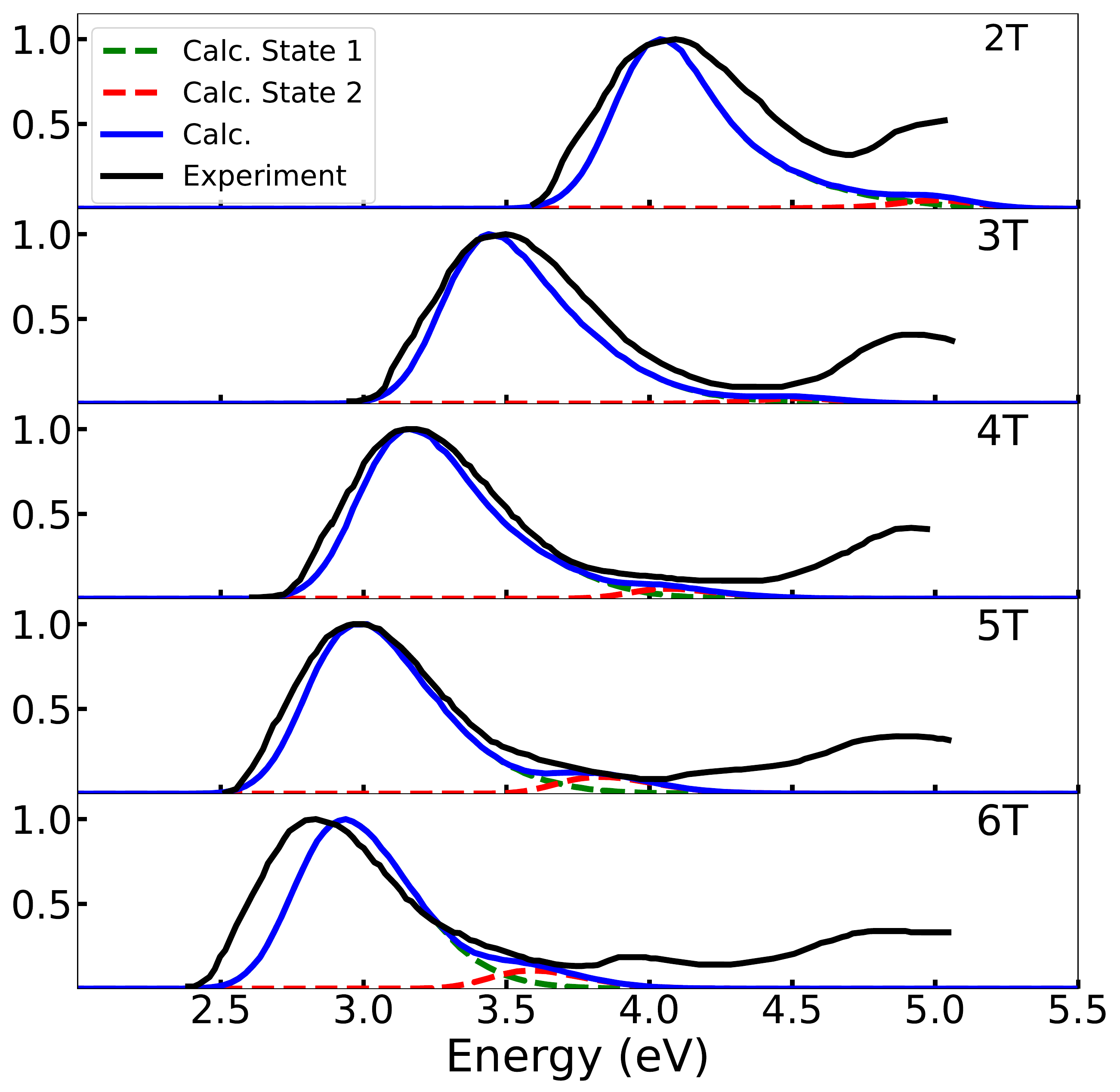}
     \caption{Experimental (black solid lines) and calculated (blue solid lines) absorption spectra of OTs in dichloromethane. The contributions from the first and second excited states to the absorption spectra are shown as green and red dashed lines, respectively. All the calculated spectra are red-shifted by 0.3 eV to facilitate the comparison with experimental spectra. Both the calculated and experimental spectra are scaled to have the same peak height.}
     \label{fig:abs}
\end{figure}

The computed absorption spectra (blue solid lines) for OTs, uniformly red-shifted by 0.3eV, are displayed in Fig. \ref{fig:abs} along with experimental ones (black solid lines), and the contributions from the first and second excited states are shown in green and red, respectively. The overall agreement between calculation and experiment is reasonable: the gradual yet nonlinear red shift of the main absorption peak from 2T to 6T, the spectral line widths (except for 2T), and the long tails on the blue side of the main peaks are well reproduced by our calculations. Note that no artificial broadening or smoothing is applied in our calculations, and this good agreement with the experiment indicates that our SchNet models, trained from high-temperature gas-phase configurations, correctly capture the wide distributions of excited-state energies resulting from the underlying structural heterogeneity of OTs in solutions. The long tail on the blue side of the main peak may be attributed to the configurations of OTs with substantial kinks and twists, and higher excited states, mostly the second excited state. Given the one million configurations used in the spectral calculation for each OT, it would be a computationally daunting task to directly apply quantum chemistry methods, such as TDDFT, and DNN models trained on quantum chemical results would be invaluable for spectral calculations, in particular for nonlinear spectroscopy. 

There are two apparent discrepancies between our calculations and experiment for all OTs. One is that our (un-shifted) calculated spectra consistently overestimate the peak positions by about 0.3-0.4eV compared to experiments. This is presumably inherited from the error associated with TDDFT against which our SchNet models were trained, as discussed in Section II of SI. The other discrepancy is that the experimental absorption spectra have extra peaks or shoulders at higher energies than the main peaks, which are originated from higher excited states beyond the first two states. We verified that if we include the first five excited states in our calculations for 2T, the high-energy peak shows up as shown in Fig. S1 of SI, although the energy spacing between the two peaks is not quantitatively reproduced. It is unclear why our calculations underestimate the width of the main peak in the spectrum of 2T, but if the predicted energy spacing between excited states is smaller, the agreement with experiment may be improved for 2T. Note that these discrepancies only suggest the limitations of TDDFT we used to train our DNN models, but they do not undermine the effectiveness of SchNet in predicting the excited-state properties at the targeted theory level, i.e., TDDFT in our case. It is also worth pointing out that a similar level of agreement between calculated and experimental spectra may be achieved even with a training set of 5000 randomly chosen configurations, as we show in Fig. S2 of SI.

\section{Conclusions and Outlook}\label{sec:conclusion}
Understanding the effect of energetic disorder on OSCs is crucial to the design of high-performance organic semiconducting devices. However, accurate simulation of such dependence requires repetitive high-level electronic structure calculations over a large number of molecular configurations, a computationally challenging task for many OSC applications. In this paper, we use deep learning methods to address this challenge and show that state-of-the-art DNNs are capable of predicting electronic properties of OSCs with an accuracy comparable to TDDFT. We demonstrate that DNNs retain their superior performance in predicting the electronic properties of OSCs even as molecule size increases. We find SchNet to be the best performing model among the four DNNs tested, achieving MAEs significantly below 0.1eV even with as few as 5000 training data. SchNet also consistently outperforms shallow feed-forward neural networks, especially in difficult cases with large molecules or limited training data. By analyzing the SchNet predicted results, we show that some 6T configurations possess strong charge-transfer character, and the transition dipole moments of these configurations could be difficult to predict using ML methods. This finding demonstrates that careful analysis of ML results, especially cases where ML methods fail, could provide physical insights into the problems of interest.

Finally, we use SchNet trained on data from 1000K gas-phase OTs to compute the absorption spectra of OTs in solution phase at room temperature and observe a good agreement with experimental results, demonstrating the transferability of DNNs in simulating OSCs under different conditions. We further show that as few as several thousand data are sufficient to train a DNN model to obtain accurate absorption spectra, greatly reducing the number of quantum chemical calculations needed in the atomistic simulations of spectral line shapes, which often require tens to hundreds of thousand electronic structure calculations if no empirical parameters are introduced. Such reduction of computational time opens the door to routine calculation of electronic spectroscopy. For example, generating 5000 training data of 6T is estimated to require only about 7500 CPU hours with Intel Core i9-9900 processors. Furthermore, the protocol for calculating absorption spectra described in this work could also be applied to other spectroscopy, such as vibrational spectroscopy.

Looking forward, DNNs could be useful for other OSC applications not yet explored in this work. For example, computing the transport coefficients of exciton/charge in extended semiconducting materials would require the evaluations of inter-molecular couplings in addition to (local) excitation energies, the computational cost of such calculations nominally scales as $O(N^2)$ where $N$ is the number of molecules. This is clearly beyond the limits of most computational resources unless drastic approximations are made~\cite{Shi2018, Lee2019}, and ML approaches could again be a potential solution to this computationally challenging task. On the other hand, generating sufficient training data might not be feasible for larger OSC molecules, in particular polymeric OSCs. In such cases, transfer learning method could be useful by pre-training ML models with data from other sources where data are abundant. In fact, there are already several recent successful demonstrations of transfer learning in chemistry and materials science applications with limited data~\cite{Wu2019, Smith2019, Yamada2019}, and we expect the performance of DNNs could be further enhanced by transfer learning and this will be a topic of future work.

\section{Associated Content}
{\bf Supporting Information.} The implementation details of DNNs, comparison of TDDFT results against a higher-level electronic structure method, and more discussions on absorption spectral simulations for OTs in dichloromethane.

\section{Acknowledgements}
L.S. acknowledges the support from the University of California Merced start-up funding.
This research was partially supported by grants from the National Natural Science Foundation of China to Q.L. (Grants No. 61922073, 61672483). Q.L. gratefully acknowledges the support of the Young Elite Scientist Sponsorship Program of CAST and the Youth Innovation Promotion Association of CAS (No. 2014299).

\bibliographystyle{naturemag_noURL}
\bibliography{MyCollection}

\providecommand{\latin}[1]{#1}
\makeatletter
\providecommand{\doi}
  {\begingroup\let\do\@makeother\dospecials
  \catcode`\{=1 \catcode`\}=2 \doi@aux}
\providecommand{\doi@aux}[1]{\endgroup\texttt{#1}}
\makeatother
\providecommand*\mcitethebibliography{\thebibliography}
\csname @ifundefined\endcsname{endmcitethebibliography}
  {\let\endmcitethebibliography\endthebibliography}{}
\begin{mcitethebibliography}{113}
\providecommand*\natexlab[1]{#1}
\providecommand*\mciteSetBstSublistMode[1]{}
\providecommand*\mciteSetBstMaxWidthForm[2]{}
\providecommand*\mciteBstWouldAddEndPuncttrue
  {\def\EndOfBibitem{\unskip.}}
\providecommand*\mciteBstWouldAddEndPunctfalse
  {\let\EndOfBibitem\relax}
\providecommand*\mciteSetBstMidEndSepPunct[3]{}
\providecommand*\mciteSetBstSublistLabelBeginEnd[3]{}
\providecommand*\EndOfBibitem{}
\mciteSetBstSublistMode{f}
\mciteSetBstMaxWidthForm{subitem}{(\alph{mcitesubitemcount})}
\mciteSetBstSublistLabelBeginEnd
  {\mcitemaxwidthsubitemform\space}
  {\relax}
  {\relax}

\bibitem[Ostroverkhova(2016)]{Ostroverkhova2016}
Ostroverkhova,~O. {Organic Optoelectronic Materials: Mechanisms and
  Applications}. \emph{Chemical Reviews} \textbf{2016}, \emph{116},
  13279--13412\relax
\mciteBstWouldAddEndPuncttrue
\mciteSetBstMidEndSepPunct{\mcitedefaultmidpunct}
{\mcitedefaultendpunct}{\mcitedefaultseppunct}\relax
\EndOfBibitem
\bibitem[Hains \latin{et~al.}(2010)Hains, Liang, Woodhouse, and
  Gregg]{Hains2010}
Hains,~A.~W.; Liang,~Z.; Woodhouse,~M.~A.; Gregg,~B.~A. {Molecular
  Semiconductors in Organic Photovoltaic Cells}. \emph{Chemical Reviews}
  \textbf{2010}, \emph{110}, 6689--6735\relax
\mciteBstWouldAddEndPuncttrue
\mciteSetBstMidEndSepPunct{\mcitedefaultmidpunct}
{\mcitedefaultendpunct}{\mcitedefaultseppunct}\relax
\EndOfBibitem
\bibitem[Myers and Xue(2012)Myers, and Xue]{Myers2012}
Myers,~J.~D.; Xue,~J. {Organic Semiconductors and their Applications in
  Photovoltaic Devices}. \emph{Polymer Reviews} \textbf{2012}, \emph{52},
  1--37\relax
\mciteBstWouldAddEndPuncttrue
\mciteSetBstMidEndSepPunct{\mcitedefaultmidpunct}
{\mcitedefaultendpunct}{\mcitedefaultseppunct}\relax
\EndOfBibitem
\bibitem[Lu \latin{et~al.}(2015)Lu, Zheng, Wu, Schneider, Zhao, and Yu]{Lu2015}
Lu,~L.; Zheng,~T.; Wu,~Q.; Schneider,~A.~M.; Zhao,~D.; Yu,~L. {Recent Advances
  in Bulk Heterojunction Polymer Solar Cells}. \emph{Chemical Reviews}
  \textbf{2015}, \emph{115}, 12666--12731\relax
\mciteBstWouldAddEndPuncttrue
\mciteSetBstMidEndSepPunct{\mcitedefaultmidpunct}
{\mcitedefaultendpunct}{\mcitedefaultseppunct}\relax
\EndOfBibitem
\bibitem[Hedley \latin{et~al.}(2017)Hedley, Ruseckas, and Samuel]{Hedley2017}
Hedley,~G.~J.; Ruseckas,~A.; Samuel,~I. D.~W. {Light Harvesting for Organic
  Photovoltaics}. \emph{Chemical Reviews} \textbf{2017}, \emph{117},
  796--837\relax
\mciteBstWouldAddEndPuncttrue
\mciteSetBstMidEndSepPunct{\mcitedefaultmidpunct}
{\mcitedefaultendpunct}{\mcitedefaultseppunct}\relax
\EndOfBibitem
\bibitem[Minaev \latin{et~al.}(2014)Minaev, Baryshnikov, and Agren]{Minaev2014}
Minaev,~B.; Baryshnikov,~G.; Agren,~H. {Principles of phosphorescent organic
  light emitting devices}. \emph{Phys. Chem. Chem. Phys.} \textbf{2014},
  \emph{16}, 1719--1758\relax
\mciteBstWouldAddEndPuncttrue
\mciteSetBstMidEndSepPunct{\mcitedefaultmidpunct}
{\mcitedefaultendpunct}{\mcitedefaultseppunct}\relax
\EndOfBibitem
\bibitem[Xu \latin{et~al.}(2016)Xu, Li, and Tang]{Xu2016}
Xu,~R.-P.; Li,~Y.-Q.; Tang,~J.-X. {Recent advances in flexible organic
  light-emitting diodes}. \emph{Journal of Materials Chemistry C}
  \textbf{2016}, \emph{4}, 9116--9142\relax
\mciteBstWouldAddEndPuncttrue
\mciteSetBstMidEndSepPunct{\mcitedefaultmidpunct}
{\mcitedefaultendpunct}{\mcitedefaultseppunct}\relax
\EndOfBibitem
\bibitem[Sirringhaus(2014)]{Sirringhaus2014}
Sirringhaus,~H. {25th Anniversary Article: Organic Field-Effect Transistors:
  The Path Beyond Amorphous Silicon}. \emph{Advanced Materials} \textbf{2014},
  \emph{26}, 1319--1335\relax
\mciteBstWouldAddEndPuncttrue
\mciteSetBstMidEndSepPunct{\mcitedefaultmidpunct}
{\mcitedefaultendpunct}{\mcitedefaultseppunct}\relax
\EndOfBibitem
\bibitem[Baeg \latin{et~al.}(2013)Baeg, Binda, Natali, Caironi, and
  Noh]{Baeg2013}
Baeg,~K.-J.; Binda,~M.; Natali,~D.; Caironi,~M.; Noh,~Y.-Y. {Organic Light
  Detectors: Photodiodes and Phototransistors}. \emph{Advanced Materials}
  \textbf{2013}, \emph{25}, 4267--4295\relax
\mciteBstWouldAddEndPuncttrue
\mciteSetBstMidEndSepPunct{\mcitedefaultmidpunct}
{\mcitedefaultendpunct}{\mcitedefaultseppunct}\relax
\EndOfBibitem
\bibitem[Moul{\'{e}} and Meerholz(2008)Moul{\'{e}}, and Meerholz]{Moule2008}
Moul{\'{e}},~A.~J.; Meerholz,~K. {Controlling morphology in polymer-fullerene
  mixtures}. \emph{Advanced Materials} \textbf{2008}, \emph{20}, 240--245\relax
\mciteBstWouldAddEndPuncttrue
\mciteSetBstMidEndSepPunct{\mcitedefaultmidpunct}
{\mcitedefaultendpunct}{\mcitedefaultseppunct}\relax
\EndOfBibitem
\bibitem[McMahon and Troisi(2010)McMahon, and Troisi]{McMahon2010}
McMahon,~D.~P.; Troisi,~A. {Organic Semiconductors: Impact of Disorder at
  Different Timescales}. \emph{ChemPhysChem} \textbf{2010}, \emph{11},
  2067--2074\relax
\mciteBstWouldAddEndPuncttrue
\mciteSetBstMidEndSepPunct{\mcitedefaultmidpunct}
{\mcitedefaultendpunct}{\mcitedefaultseppunct}\relax
\EndOfBibitem
\bibitem[Br{\'{e}}das \latin{et~al.}(2009)Br{\'{e}}das, Norton, Cornil, and
  Coropceanu]{Bredas2009}
Br{\'{e}}das,~J.-L.; Norton,~J.~E.; Cornil,~J.; Coropceanu,~V. {Molecular
  Understanding of Organic Solar Cells: The Challenges}. \emph{Accounts of
  Chemical Research} \textbf{2009}, \emph{42}, 1691--1699\relax
\mciteBstWouldAddEndPuncttrue
\mciteSetBstMidEndSepPunct{\mcitedefaultmidpunct}
{\mcitedefaultendpunct}{\mcitedefaultseppunct}\relax
\EndOfBibitem
\bibitem[Troisi(2011)]{Troisi2011}
Troisi,~A. {Charge transport in high mobility molecular semiconductors:
  classical models and new theories}. \emph{Chemical Society Reviews}
  \textbf{2011}, \emph{40}, 2347\relax
\mciteBstWouldAddEndPuncttrue
\mciteSetBstMidEndSepPunct{\mcitedefaultmidpunct}
{\mcitedefaultendpunct}{\mcitedefaultseppunct}\relax
\EndOfBibitem
\bibitem[Zhugayevych and Tretiak(2015)Zhugayevych, and
  Tretiak]{Zhugayevych2014}
Zhugayevych,~A.; Tretiak,~S. {Theoretical Description of Structural and
  Electronic Properties of Organic Photovoltaic Materials}. \emph{Annual Review
  of Physical Chemistry} \textbf{2015}, \emph{66}, 305--330\relax
\mciteBstWouldAddEndPuncttrue
\mciteSetBstMidEndSepPunct{\mcitedefaultmidpunct}
{\mcitedefaultendpunct}{\mcitedefaultseppunct}\relax
\EndOfBibitem
\bibitem[Salzner and Aydin(2011)Salzner, and Aydin]{Salzner2011}
Salzner,~U.; Aydin,~A. {Improved Prediction of Properties of $\pi$-Conjugated
  Oligomers with Range-Separated Hybrid Density Functionals}. \emph{Journal of
  Chemical Theory and Computation} \textbf{2011}, \emph{7}, 2568--2583\relax
\mciteBstWouldAddEndPuncttrue
\mciteSetBstMidEndSepPunct{\mcitedefaultmidpunct}
{\mcitedefaultendpunct}{\mcitedefaultseppunct}\relax
\EndOfBibitem
\bibitem[Jacquemin \latin{et~al.}(2015)Jacquemin, Duchemin, and
  Blase]{Jacquemin2015}
Jacquemin,~D.; Duchemin,~I.; Blase,~X. {0-0 Energies Using Hybrid Schemes:
  Benchmarks of TD-DFT, CIS(D), ADC(2), CC2, and BSE/ GW formalisms for 80
  Real-Life Compounds}. \emph{Journal of Chemical Theory and Computation}
  \textbf{2015}, \emph{11}, 5340--5359\relax
\mciteBstWouldAddEndPuncttrue
\mciteSetBstMidEndSepPunct{\mcitedefaultmidpunct}
{\mcitedefaultendpunct}{\mcitedefaultseppunct}\relax
\EndOfBibitem
\bibitem[Faber \latin{et~al.}(2014)Faber, Boulanger, Attaccalite, Duchemin, and
  Blase]{Faber2014}
Faber,~C.; Boulanger,~P.; Attaccalite,~C.; Duchemin,~I.; Blase,~X. {Excited
  states properties of organic molecules: from density functional theory to the
  GW and Bethe–Salpeter Green's function formalisms}. \emph{Philosophical
  Transactions of the Royal Society A: Mathematical, Physical and Engineering
  Sciences} \textbf{2014}, \emph{372}, 20130271\relax
\mciteBstWouldAddEndPuncttrue
\mciteSetBstMidEndSepPunct{\mcitedefaultmidpunct}
{\mcitedefaultendpunct}{\mcitedefaultseppunct}\relax
\EndOfBibitem
\bibitem[K{\"{u}}mmel(2017)]{Kummel2017}
K{\"{u}}mmel,~S. {Charge-Transfer Excitations: A Challenge for Time-Dependent
  Density Functional Theory That Has Been Met}. \emph{Advanced Energy
  Materials} \textbf{2017}, \emph{7}, 1700440\relax
\mciteBstWouldAddEndPuncttrue
\mciteSetBstMidEndSepPunct{\mcitedefaultmidpunct}
{\mcitedefaultendpunct}{\mcitedefaultseppunct}\relax
\EndOfBibitem
\bibitem[K{\"{o}}rzd{\"{o}}rfer and Br{\'{e}}das(2014)K{\"{o}}rzd{\"{o}}rfer,
  and Br{\'{e}}das]{Korzdorfer2014}
K{\"{o}}rzd{\"{o}}rfer,~T.; Br{\'{e}}das,~J.-L. {Organic Electronic Materials:
  Recent Advances in the DFT Description of the Ground and Excited States Using
  Tuned Range-Separated Hybrid Functionals}. \emph{Accounts of Chemical
  Research} \textbf{2014}, \emph{47}, 3284--3291\relax
\mciteBstWouldAddEndPuncttrue
\mciteSetBstMidEndSepPunct{\mcitedefaultmidpunct}
{\mcitedefaultendpunct}{\mcitedefaultseppunct}\relax
\EndOfBibitem
\bibitem[Feron \latin{et~al.}(2012)Feron, Zhou, Belcher, and
  Dastoor]{Feron2012a}
Feron,~K.; Zhou,~X.; Belcher,~W.~J.; Dastoor,~P.~C. {Exciton transport in
  organic semiconductors: F{\"{o}}rster resonance energy transfer compared with
  a simple random walk}. \emph{Journal of Applied Physics} \textbf{2012},
  \emph{111}, 044510\relax
\mciteBstWouldAddEndPuncttrue
\mciteSetBstMidEndSepPunct{\mcitedefaultmidpunct}
{\mcitedefaultendpunct}{\mcitedefaultseppunct}\relax
\EndOfBibitem
\bibitem[Feron \latin{et~al.}(2012)Feron, Belcher, Fell, and
  Dastoor]{Feron2012}
Feron,~K.; Belcher,~W.; Fell,~C.; Dastoor,~P. {Organic Solar Cells:
  Understanding the Role of F{\"{o}}rster Resonance Energy Transfer}.
  \emph{International Journal of Molecular Sciences} \textbf{2012}, \emph{13},
  17019--17047\relax
\mciteBstWouldAddEndPuncttrue
\mciteSetBstMidEndSepPunct{\mcitedefaultmidpunct}
{\mcitedefaultendpunct}{\mcitedefaultseppunct}\relax
\EndOfBibitem
\bibitem[Li \latin{et~al.}(2013)Li, Malinin, Tretiak, and Chernyak]{Li2013}
Li,~H.; Malinin,~S.~V.; Tretiak,~S.; Chernyak,~V.~Y. {Effective tight-binding
  models for excitons in branched conjugated molecules}. \emph{The Journal of
  Chemical Physics} \textbf{2013}, \emph{139}, 064109\relax
\mciteBstWouldAddEndPuncttrue
\mciteSetBstMidEndSepPunct{\mcitedefaultmidpunct}
{\mcitedefaultendpunct}{\mcitedefaultseppunct}\relax
\EndOfBibitem
\bibitem[Tapping \latin{et~al.}(2015)Tapping, Clafton, Schwarz, Kee, and
  Huang]{Tapping2015}
Tapping,~P.~C.; Clafton,~S.~N.; Schwarz,~K.~N.; Kee,~T.~W.; Huang,~D.~M.
  {Molecular-Level Details of Morphology-Dependent Exciton Migration in
  Poly(3-hexylthiophene) Nanostructures}. \emph{The Journal of Physical
  Chemistry C} \textbf{2015}, \emph{119}, 7047--7059\relax
\mciteBstWouldAddEndPuncttrue
\mciteSetBstMidEndSepPunct{\mcitedefaultmidpunct}
{\mcitedefaultendpunct}{\mcitedefaultseppunct}\relax
\EndOfBibitem
\bibitem[B{\"{a}}ssler(1993)]{Bassler1993}
B{\"{a}}ssler,~H. {Charge Transport in Disordered Organic Photoconductors a
  Monte Carlo Simulation Study}. \emph{Physica Status Solidi (b)}
  \textbf{1993}, \emph{175}, 15--56\relax
\mciteBstWouldAddEndPuncttrue
\mciteSetBstMidEndSepPunct{\mcitedefaultmidpunct}
{\mcitedefaultendpunct}{\mcitedefaultseppunct}\relax
\EndOfBibitem
\bibitem[Shi \latin{et~al.}(2017)Shi, Lee, and Willard]{Shi2017}
Shi,~L.; Lee,~C.~K.; Willard,~A.~P. {The Enhancement of Interfacial Exciton
  Dissociation by Energetic Disorder Is a Nonequilibrium Effect}. \emph{ACS
  Central Science} \textbf{2017}, \emph{3}, 1262--1270\relax
\mciteBstWouldAddEndPuncttrue
\mciteSetBstMidEndSepPunct{\mcitedefaultmidpunct}
{\mcitedefaultendpunct}{\mcitedefaultseppunct}\relax
\EndOfBibitem
\bibitem[Lee \latin{et~al.}(2016)Lee, Shi, and Willard]{Lee2016}
Lee,~C.~K.; Shi,~L.; Willard,~A.~P. {A Model of Charge-Transfer Excitons:
  Diffusion, Spin Dynamics, and Magnetic Field Effects}. \emph{The Journal of
  Physical Chemistry Letters} \textbf{2016}, \emph{7}, 2246--2251\relax
\mciteBstWouldAddEndPuncttrue
\mciteSetBstMidEndSepPunct{\mcitedefaultmidpunct}
{\mcitedefaultendpunct}{\mcitedefaultseppunct}\relax
\EndOfBibitem
\bibitem[Ramakrishnan \latin{et~al.}(2015)Ramakrishnan, Dral, Rupp, and von
  Lilienfeld]{Ramakrishnan2015}
Ramakrishnan,~R.; Dral,~P.~O.; Rupp,~M.; von Lilienfeld,~O.~A. {Big Data Meets
  Quantum Chemistry Approximations: The $\Delta$-Machine Learning Approach}.
  \emph{Journal of Chemical Theory and Computation} \textbf{2015}, \emph{11},
  2087--2096\relax
\mciteBstWouldAddEndPuncttrue
\mciteSetBstMidEndSepPunct{\mcitedefaultmidpunct}
{\mcitedefaultendpunct}{\mcitedefaultseppunct}\relax
\EndOfBibitem
\bibitem[Welborn \latin{et~al.}(2018)Welborn, Cheng, and Miller]{Welborn2018}
Welborn,~M.; Cheng,~L.; Miller,~T.~F. {Transferability in Machine Learning for
  Electronic Structure via the Molecular Orbital Basis}. \emph{Journal of
  Chemical Theory and Computation} \textbf{2018}, \emph{14}, 4772--4779\relax
\mciteBstWouldAddEndPuncttrue
\mciteSetBstMidEndSepPunct{\mcitedefaultmidpunct}
{\mcitedefaultendpunct}{\mcitedefaultseppunct}\relax
\EndOfBibitem
\bibitem[Zaspel \latin{et~al.}(2019)Zaspel, Huang, Harbrecht, and von
  Lilienfeld]{Zaspel2018}
Zaspel,~P.; Huang,~B.; Harbrecht,~H.; von Lilienfeld,~O.~A. {Boosting Quantum
  Machine Learning Models with a Multilevel Combination Technique: Pople
  Diagrams Revisited}. \emph{Journal of Chemical Theory and Computation}
  \textbf{2019}, \emph{15}, 1546--1559\relax
\mciteBstWouldAddEndPuncttrue
\mciteSetBstMidEndSepPunct{\mcitedefaultmidpunct}
{\mcitedefaultendpunct}{\mcitedefaultseppunct}\relax
\EndOfBibitem
\bibitem[Cheng \latin{et~al.}(2019)Cheng, Welborn, Christensen, and
  Miller]{Cheng2019}
Cheng,~L.; Welborn,~M.; Christensen,~A.~S.; Miller,~T.~F. {A universal density
  matrix functional from molecular orbital-based machine learning:
  Transferability across organic molecules}. \emph{The Journal of Chemical
  Physics} \textbf{2019}, \emph{150}, 131103\relax
\mciteBstWouldAddEndPuncttrue
\mciteSetBstMidEndSepPunct{\mcitedefaultmidpunct}
{\mcitedefaultendpunct}{\mcitedefaultseppunct}\relax
\EndOfBibitem
\bibitem[Brockherde \latin{et~al.}(2017)Brockherde, Vogt, Li, Tuckerman, Burke,
  and M{\"{u}}ller]{Brockherde2017}
Brockherde,~F.; Vogt,~L.; Li,~L.; Tuckerman,~M.~E.; Burke,~K.;
  M{\"{u}}ller,~K.-R. {Bypassing the Kohn-Sham equations with machine
  learning}. \emph{Nature Communications} \textbf{2017}, \emph{8}, 872\relax
\mciteBstWouldAddEndPuncttrue
\mciteSetBstMidEndSepPunct{\mcitedefaultmidpunct}
{\mcitedefaultendpunct}{\mcitedefaultseppunct}\relax
\EndOfBibitem
\bibitem[Bogojeski \latin{et~al.}(2018)Bogojeski, Brockherde, Vogt-Maranto, Li,
  Tuckerman, Burke, and M{\"{u}}ller]{Bogojeski2018}
Bogojeski,~M.; Brockherde,~F.; Vogt-Maranto,~L.; Li,~L.; Tuckerman,~M.~E.;
  Burke,~K.; M{\"{u}}ller,~K.-R. {Efficient prediction of 3D electron densities
  using machine learning}. \textbf{2018}, arXiv:1811.06255\relax
\mciteBstWouldAddEndPuncttrue
\mciteSetBstMidEndSepPunct{\mcitedefaultmidpunct}
{\mcitedefaultendpunct}{\mcitedefaultseppunct}\relax
\EndOfBibitem
\bibitem[Grisafi \latin{et~al.}(2019)Grisafi, Fabrizio, Meyer, Wilkins,
  Corminboeuf, and Ceriotti]{Grisafi2018}
Grisafi,~A.; Fabrizio,~A.; Meyer,~B.; Wilkins,~D.~M.; Corminboeuf,~C.;
  Ceriotti,~M. {Transferable Machine-Learning Model of the Electron Density}.
  \emph{ACS Central Science} \textbf{2019}, \emph{5}, 57--64\relax
\mciteBstWouldAddEndPuncttrue
\mciteSetBstMidEndSepPunct{\mcitedefaultmidpunct}
{\mcitedefaultendpunct}{\mcitedefaultseppunct}\relax
\EndOfBibitem
\bibitem[Fabrizio \latin{et~al.}(2019)Fabrizio, Grisafi, Meyer, Ceriotti, and
  Corminboeuf]{Fabrizio2019}
Fabrizio,~A.; Grisafi,~A.; Meyer,~B.; Ceriotti,~M.; Corminboeuf,~C. {Electron
  density learning of non-covalent systems}. \emph{Chemical Science}
  \textbf{2019}, \relax
\mciteBstWouldAddEndPunctfalse
\mciteSetBstMidEndSepPunct{\mcitedefaultmidpunct}
{}{\mcitedefaultseppunct}\relax
\EndOfBibitem
\bibitem[Ryczko \latin{et~al.}(2019)Ryczko, Strubbe, and Tamblyn]{Ryczko2019}
Ryczko,~K.; Strubbe,~D.~A.; Tamblyn,~I. {Deep learning and density-functional
  theory}. \emph{Physical Review A} \textbf{2019}, \emph{100}, 022512\relax
\mciteBstWouldAddEndPuncttrue
\mciteSetBstMidEndSepPunct{\mcitedefaultmidpunct}
{\mcitedefaultendpunct}{\mcitedefaultseppunct}\relax
\EndOfBibitem
\bibitem[Huan \latin{et~al.}(2017)Huan, Batra, Chapman, Krishnan, Chen, and
  Ramprasad]{Huan2017}
Huan,~T.~D.; Batra,~R.; Chapman,~J.; Krishnan,~S.; Chen,~L.; Ramprasad,~R. {A
  universal strategy for the creation of machine learning-based atomistic force
  fields}. \emph{npj Computational Materials} \textbf{2017}, \emph{3}, 37\relax
\mciteBstWouldAddEndPuncttrue
\mciteSetBstMidEndSepPunct{\mcitedefaultmidpunct}
{\mcitedefaultendpunct}{\mcitedefaultseppunct}\relax
\EndOfBibitem
\bibitem[Smith \latin{et~al.}(2017)Smith, Isayev, and Roitberg]{Smith2017}
Smith,~J.~S.; Isayev,~O.; Roitberg,~A.~E. {ANI-1: an extensible neural network
  potential with DFT accuracy at force field computational cost}.
  \emph{Chemical Science} \textbf{2017}, \emph{8}, 3192--3203\relax
\mciteBstWouldAddEndPuncttrue
\mciteSetBstMidEndSepPunct{\mcitedefaultmidpunct}
{\mcitedefaultendpunct}{\mcitedefaultseppunct}\relax
\EndOfBibitem
\bibitem[Botu \latin{et~al.}(2017)Botu, Batra, Chapman, and
  Ramprasad]{Botu2016}
Botu,~V.; Batra,~R.; Chapman,~J.; Ramprasad,~R. {Machine Learning Force Fields:
  Construction, Validation, and Outlook}. \emph{The Journal of Physical
  Chemistry C} \textbf{2017}, \emph{121}, 511--522\relax
\mciteBstWouldAddEndPuncttrue
\mciteSetBstMidEndSepPunct{\mcitedefaultmidpunct}
{\mcitedefaultendpunct}{\mcitedefaultseppunct}\relax
\EndOfBibitem
\bibitem[Chmiela \latin{et~al.}(2017)Chmiela, Tkatchenko, Sauceda, Poltavsky,
  Sch{\"{u}}tt, and M{\"{u}}ller]{Chmiela2017}
Chmiela,~S.; Tkatchenko,~A.; Sauceda,~H.~E.; Poltavsky,~I.;
  Sch{\"{u}}tt,~K.~T.; M{\"{u}}ller,~K.-R. {Machine learning of accurate
  energy-conserving molecular force fields}. \emph{Science Advances}
  \textbf{2017}, \emph{3}, e1603015\relax
\mciteBstWouldAddEndPuncttrue
\mciteSetBstMidEndSepPunct{\mcitedefaultmidpunct}
{\mcitedefaultendpunct}{\mcitedefaultseppunct}\relax
\EndOfBibitem
\bibitem[Chmiela \latin{et~al.}(2018)Chmiela, Sauceda, M{\"{u}}ller, and
  Tkatchenko]{Chmiela2018}
Chmiela,~S.; Sauceda,~H.~E.; M{\"{u}}ller,~K.-r.; Tkatchenko,~A. {Towards exact
  molecular dynamics simulations with machine-learned force fields}.
  \emph{Nature Communications} \textbf{2018}, \emph{9}, 3887\relax
\mciteBstWouldAddEndPuncttrue
\mciteSetBstMidEndSepPunct{\mcitedefaultmidpunct}
{\mcitedefaultendpunct}{\mcitedefaultseppunct}\relax
\EndOfBibitem
\bibitem[Wang \latin{et~al.}(2019)Wang, Olsson, Wehmeyer, P{\'{e}}rez, Charron,
  de~Fabritiis, No{\'{e}}, and Clementi]{Wang2019}
Wang,~J.; Olsson,~S.; Wehmeyer,~C.; P{\'{e}}rez,~A.; Charron,~N.~E.;
  de~Fabritiis,~G.; No{\'{e}},~F.; Clementi,~C. {Machine Learning of
  Coarse-Grained Molecular Dynamics Force Fields}. \emph{ACS Central Science}
  \textbf{2019}, \emph{5}, 755−767\relax
\mciteBstWouldAddEndPuncttrue
\mciteSetBstMidEndSepPunct{\mcitedefaultmidpunct}
{\mcitedefaultendpunct}{\mcitedefaultseppunct}\relax
\EndOfBibitem
\bibitem[Jinnouchi \latin{et~al.}(2019)Jinnouchi, Karsai, and
  Kresse]{Jinnouchi2019}
Jinnouchi,~R.; Karsai,~F.; Kresse,~G. {On-the-fly machine learning force field
  generation: Application to melting points}. \emph{Physical Review B}
  \textbf{2019}, \emph{100}, 014105\relax
\mciteBstWouldAddEndPuncttrue
\mciteSetBstMidEndSepPunct{\mcitedefaultmidpunct}
{\mcitedefaultendpunct}{\mcitedefaultseppunct}\relax
\EndOfBibitem
\bibitem[Ramakrishnan \latin{et~al.}(2015)Ramakrishnan, Hartmann, Tapavicza,
  and von Lilienfeld]{Ramakrishnan2015b}
Ramakrishnan,~R.; Hartmann,~M.; Tapavicza,~E.; von Lilienfeld,~O.~A.
  {Electronic spectra from TDDFT and machine learning in chemical space}.
  \emph{The Journal of Chemical Physics} \textbf{2015}, \emph{143},
  084111\relax
\mciteBstWouldAddEndPuncttrue
\mciteSetBstMidEndSepPunct{\mcitedefaultmidpunct}
{\mcitedefaultendpunct}{\mcitedefaultseppunct}\relax
\EndOfBibitem
\bibitem[Ye \latin{et~al.}(2019)Ye, Hu, Li, Zhang, Zhong, Zhang, Luo, Mukamel,
  and Jiang]{Ye2019}
Ye,~S.; Hu,~W.; Li,~X.; Zhang,~J.; Zhong,~K.; Zhang,~G.; Luo,~Y.; Mukamel,~S.;
  Jiang,~J. {A neural network protocol for electronic excitations of N
  -methylacetamide}. \emph{Proceedings of the National Academy of Sciences}
  \textbf{2019}, \emph{116}, 201821044\relax
\mciteBstWouldAddEndPuncttrue
\mciteSetBstMidEndSepPunct{\mcitedefaultmidpunct}
{\mcitedefaultendpunct}{\mcitedefaultseppunct}\relax
\EndOfBibitem
\bibitem[Gastegger \latin{et~al.}(2017)Gastegger, Behler, and
  Marquetand]{Gastegger2017}
Gastegger,~M.; Behler,~J.; Marquetand,~P. {Machine learning molecular dynamics
  for the simulation of infrared spectra}. \emph{Chemical Science}
  \textbf{2017}, \emph{8}, 6924--6935\relax
\mciteBstWouldAddEndPuncttrue
\mciteSetBstMidEndSepPunct{\mcitedefaultmidpunct}
{\mcitedefaultendpunct}{\mcitedefaultseppunct}\relax
\EndOfBibitem
\bibitem[Ghosh \latin{et~al.}(2019)Ghosh, Stuke, Todorovi{\'{c}}, J{\o}rgensen,
  Schmidt, Vehtari, and Rinke]{Ghosh2019}
Ghosh,~K.; Stuke,~A.; Todorovi{\'{c}},~M.; J{\o}rgensen,~P.~B.; Schmidt,~M.~N.;
  Vehtari,~A.; Rinke,~P. {Deep Learning Spectroscopy: Neural Networks for
  Molecular Excitation Spectra}. \emph{Advanced Science} \textbf{2019},
  \emph{6}, 1801367\relax
\mciteBstWouldAddEndPuncttrue
\mciteSetBstMidEndSepPunct{\mcitedefaultmidpunct}
{\mcitedefaultendpunct}{\mcitedefaultseppunct}\relax
\EndOfBibitem
\bibitem[Jorissen and Gilson(2005)Jorissen, and Gilson]{Jorissen2005}
Jorissen,~R.~N.; Gilson,~M.~K. {Virtual Screening of Molecular Databases Using
  a Support Vector Machine}. \emph{Journal of Chemical Information and
  Modeling} \textbf{2005}, \emph{45}, 549--561\relax
\mciteBstWouldAddEndPuncttrue
\mciteSetBstMidEndSepPunct{\mcitedefaultmidpunct}
{\mcitedefaultendpunct}{\mcitedefaultseppunct}\relax
\EndOfBibitem
\bibitem[G{\'{o}}mez-Bombarelli \latin{et~al.}(2016)G{\'{o}}mez-Bombarelli,
  Aguilera-Iparraguirre, Hirzel, Duvenaud, Maclaurin, Blood-Forsythe, Chae,
  Einzinger, Ha, Wu, Markopoulos, Jeon, Kang, Miyazaki, Numata, Kim, Huang,
  Hong, Baldo, Adams, and Aspuru-Guzik]{Gomez-Bombarelli2016}
G{\'{o}}mez-Bombarelli,~R. \latin{et~al.}  {Design of efficient molecular
  organic light-emitting diodes by a high-throughput virtual screening and
  experimental approach}. \emph{Nature Materials} \textbf{2016}, \emph{15},
  1120--1127\relax
\mciteBstWouldAddEndPuncttrue
\mciteSetBstMidEndSepPunct{\mcitedefaultmidpunct}
{\mcitedefaultendpunct}{\mcitedefaultseppunct}\relax
\EndOfBibitem
\bibitem[G{\'{o}}mez-Bombarelli \latin{et~al.}(2018)G{\'{o}}mez-Bombarelli,
  Wei, Duvenaud, Hern{\'{a}}ndez-Lobato, S{\'{a}}nchez-Lengeling, Sheberla,
  Aguilera-Iparraguirre, Hirzel, Adams, and Aspuru-Guzik]{Gomez-Bombarelli2018}
G{\'{o}}mez-Bombarelli,~R.; Wei,~J.~N.; Duvenaud,~D.;
  Hern{\'{a}}ndez-Lobato,~J.~M.; S{\'{a}}nchez-Lengeling,~B.; Sheberla,~D.;
  Aguilera-Iparraguirre,~J.; Hirzel,~T.~D.; Adams,~R.~P.; Aspuru-Guzik,~A.
  {Automatic Chemical Design Using a Data-Driven Continuous Representation of
  Molecules}. \emph{ACS Central Science} \textbf{2018}, \emph{4},
  268--276\relax
\mciteBstWouldAddEndPuncttrue
\mciteSetBstMidEndSepPunct{\mcitedefaultmidpunct}
{\mcitedefaultendpunct}{\mcitedefaultseppunct}\relax
\EndOfBibitem
\bibitem[Wu \latin{et~al.}(2019)Wu, Pan, Chen, Long, Zhang, and Yu]{2019Wu}
Wu,~Z.; Pan,~S.; Chen,~F.; Long,~G.; Zhang,~C.; Yu,~P.~S. {A Comprehensive
  Survey on Graph Neural Networks}. \textbf{2019}, arXiv:1901.00596\relax
\mciteBstWouldAddEndPuncttrue
\mciteSetBstMidEndSepPunct{\mcitedefaultmidpunct}
{\mcitedefaultendpunct}{\mcitedefaultseppunct}\relax
\EndOfBibitem
\bibitem[Duvenaud \latin{et~al.}(2015)Duvenaud, Maclaurin, Iparraguirre,
  Bombarell, Hirzel, Aspuru-Guzik, and Adams]{Duvenaud2015}
Duvenaud,~D.~K.; Maclaurin,~D.; Iparraguirre,~J.; Bombarell,~R.; Hirzel,~T.;
  Aspuru-Guzik,~A.; Adams,~R.~P. {Convolutional Networks on Graphs for Learning
  Molecular Fingerprints}. \emph{Advances in Neural Information Processing
  Systems} \textbf{2015}, 2224--2232\relax
\mciteBstWouldAddEndPuncttrue
\mciteSetBstMidEndSepPunct{\mcitedefaultmidpunct}
{\mcitedefaultendpunct}{\mcitedefaultseppunct}\relax
\EndOfBibitem
\bibitem[Sch{\"{u}}tt \latin{et~al.}(2017)Sch{\"{u}}tt, Arbabzadah, Chmiela,
  M{\"{u}}ller, and Tkatchenko]{Schutt2017}
Sch{\"{u}}tt,~K.~T.; Arbabzadah,~F.; Chmiela,~S.; M{\"{u}}ller,~K.~R.;
  Tkatchenko,~A. {Quantum-chemical insights from deep tensor neural networks}.
  \emph{Nature Communications} \textbf{2017}, \emph{8}, 13890\relax
\mciteBstWouldAddEndPuncttrue
\mciteSetBstMidEndSepPunct{\mcitedefaultmidpunct}
{\mcitedefaultendpunct}{\mcitedefaultseppunct}\relax
\EndOfBibitem
\bibitem[Gilmer \latin{et~al.}(2017)Gilmer, Schoenholz, Riley, Vinyals, and
  Dahl]{Gilmer2017}
Gilmer,~J.; Schoenholz,~S.~S.; Riley,~P.~F.; Vinyals,~O.; Dahl,~G.~E. {Neural
  message passing for quantum chemistry}. Proceedings of the 34th International
  Conference on Machine Learning-Volume 70. 2017; pp 1263--1272\relax
\mciteBstWouldAddEndPuncttrue
\mciteSetBstMidEndSepPunct{\mcitedefaultmidpunct}
{\mcitedefaultendpunct}{\mcitedefaultseppunct}\relax
\EndOfBibitem
\bibitem[Sch{\"{u}}tt \latin{et~al.}(2018)Sch{\"{u}}tt, Sauceda, Kindermans,
  Tkatchenko, and M{\"{u}}ller]{Schutt2018}
Sch{\"{u}}tt,~K.~T.; Sauceda,~H.~E.; Kindermans,~P.-J.; Tkatchenko,~A.;
  M{\"{u}}ller,~K.-R. {SchNet - A deep learning architecture for molecules and
  materials}. \emph{The Journal of Chemical Physics} \textbf{2018}, \emph{148},
  241722\relax
\mciteBstWouldAddEndPuncttrue
\mciteSetBstMidEndSepPunct{\mcitedefaultmidpunct}
{\mcitedefaultendpunct}{\mcitedefaultseppunct}\relax
\EndOfBibitem
\bibitem[Wu \latin{et~al.}(2018)Wu, Ramsundar, Feinberg, Gomes, Geniesse,
  Pappu, Leswing, and Pande]{Wu2018}
Wu,~Z.; Ramsundar,~B.; Feinberg,~E.~N.; Gomes,~J.; Geniesse,~C.; Pappu,~A.~S.;
  Leswing,~K.; Pande,~V. {MoleculeNet: a benchmark for molecular machine
  learning}. Chemical Science. 2018; pp 513--530\relax
\mciteBstWouldAddEndPuncttrue
\mciteSetBstMidEndSepPunct{\mcitedefaultmidpunct}
{\mcitedefaultendpunct}{\mcitedefaultseppunct}\relax
\EndOfBibitem
\bibitem[Unke and Meuwly(2019)Unke, and Meuwly]{Unke2019}
Unke,~O.~T.; Meuwly,~M. {PhysNet: A Neural Network for Predicting Energies,
  Forces, Dipole Moments, and Partial Charges}. \emph{Journal of Chemical
  Theory and Computation} \textbf{2019}, \emph{15}, 3678--3693\relax
\mciteBstWouldAddEndPuncttrue
\mciteSetBstMidEndSepPunct{\mcitedefaultmidpunct}
{\mcitedefaultendpunct}{\mcitedefaultseppunct}\relax
\EndOfBibitem
\bibitem[Sch{\"{u}}tt \latin{et~al.}(2019)Sch{\"{u}}tt, Gastegger, Tkatchenko,
  M{\"{u}}ller, and Maurer]{Schutt2019}
Sch{\"{u}}tt,~K.~T.; Gastegger,~M.; Tkatchenko,~A.; M{\"{u}}ller,~K.~R.;
  Maurer,~R.~J. {Unifying machine learning and quantum chemistry -- a deep
  neural network for molecular wavefunctions}. \textbf{2019},
  arXiv:1906.10033\relax
\mciteBstWouldAddEndPuncttrue
\mciteSetBstMidEndSepPunct{\mcitedefaultmidpunct}
{\mcitedefaultendpunct}{\mcitedefaultseppunct}\relax
\EndOfBibitem
\bibitem[Lu \latin{et~al.}(2019)Lu, Liu, Wang, Huang, Lin, and He]{Lu2019a}
Lu,~C.; Liu,~Q.; Wang,~C.; Huang,~Z.; Lin,~P.; He,~L. {Molecular Property
  Prediction: A Multilevel Quantum Interactions Modeling Perspective}.
  \emph{Proceedings of the AAAI Conference on Artificial Intelligence}
  \textbf{2019}, \emph{33}, 1052--1060\relax
\mciteBstWouldAddEndPuncttrue
\mciteSetBstMidEndSepPunct{\mcitedefaultmidpunct}
{\mcitedefaultendpunct}{\mcitedefaultseppunct}\relax
\EndOfBibitem
\bibitem[Chen \latin{et~al.}(2019)Chen, Chen, Hsieh, Lee, Liao, Liao, Liu, Qiu,
  Sun, Tang, Zemel, and Zhang]{Chen2019}
Chen,~G.; Chen,~P.; Hsieh,~C.-Y.; Lee,~C.-K.; Liao,~B.; Liao,~R.; Liu,~W.;
  Qiu,~J.; Sun,~Q.; Tang,~J.; Zemel,~R.; Zhang,~S. {Alchemy: A Quantum
  Chemistry Dataset for Benchmarking AI Models}. \textbf{2019},
  arXiv:1906.09427\relax
\mciteBstWouldAddEndPuncttrue
\mciteSetBstMidEndSepPunct{\mcitedefaultmidpunct}
{\mcitedefaultendpunct}{\mcitedefaultseppunct}\relax
\EndOfBibitem
\bibitem[Rupp \latin{et~al.}(2012)Rupp, Tkatchenko, M{\"{u}}ller, and von
  Lilienfeld]{Rupp2012}
Rupp,~M.; Tkatchenko,~A.; M{\"{u}}ller,~K.-R.; von Lilienfeld,~O.~A. {Fast and
  Accurate Modeling of Molecular Atomization Energies with Machine Learning}.
  \emph{Physical Review Letters} \textbf{2012}, \emph{108}, 058301\relax
\mciteBstWouldAddEndPuncttrue
\mciteSetBstMidEndSepPunct{\mcitedefaultmidpunct}
{\mcitedefaultendpunct}{\mcitedefaultseppunct}\relax
\EndOfBibitem
\bibitem[Hansen \latin{et~al.}(2013)Hansen, Montavon, Biegler, Fazli, Rupp,
  Scheffler, von Lilienfeld, Tkatchenko, and M{\"{u}}ller]{Hansen2013}
Hansen,~K.; Montavon,~G.; Biegler,~F.; Fazli,~S.; Rupp,~M.; Scheffler,~M.; von
  Lilienfeld,~O.~A.; Tkatchenko,~A.; M{\"{u}}ller,~K.-R. {Assessment and
  Validation of Machine Learning Methods for Predicting Molecular Atomization
  Energies}. \emph{Journal of Chemical Theory and Computation} \textbf{2013},
  \emph{9}, 3404--3419\relax
\mciteBstWouldAddEndPuncttrue
\mciteSetBstMidEndSepPunct{\mcitedefaultmidpunct}
{\mcitedefaultendpunct}{\mcitedefaultseppunct}\relax
\EndOfBibitem
\bibitem[Ramakrishnan \latin{et~al.}(2014)Ramakrishnan, Dral, Rupp, and von
  Lilienfeld]{Ramakrishnan2014}
Ramakrishnan,~R.; Dral,~P.~O.; Rupp,~M.; von Lilienfeld,~O.~A. {Quantum
  chemistry structures and properties of 134 kilo molecules}. \emph{Scientific
  Data} \textbf{2014}, \emph{1}, 140022\relax
\mciteBstWouldAddEndPuncttrue
\mciteSetBstMidEndSepPunct{\mcitedefaultmidpunct}
{\mcitedefaultendpunct}{\mcitedefaultseppunct}\relax
\EndOfBibitem
\bibitem[{St. John} \latin{et~al.}(2019){St. John}, Phillips, Kemper, Wilson,
  Guan, Crowley, Nimlos, and Larsen]{St.John2019}
{St. John},~P.~C.; Phillips,~C.; Kemper,~T.~W.; Wilson,~A.~N.; Guan,~Y.;
  Crowley,~M.~F.; Nimlos,~M.~R.; Larsen,~R.~E. {Message-passing neural networks
  for high-throughput polymer screening}. \emph{The Journal of Chemical
  Physics} \textbf{2019}, \emph{150}, 234111\relax
\mciteBstWouldAddEndPuncttrue
\mciteSetBstMidEndSepPunct{\mcitedefaultmidpunct}
{\mcitedefaultendpunct}{\mcitedefaultseppunct}\relax
\EndOfBibitem
\bibitem[Fulde(1995)]{Fulde1995}
Fulde,~P. \emph{Electron Correlations in Molecules and Solids}; Springer Berlin
  Heidelberg: Berlin, Heidelberg, 1995; pp 107--128\relax
\mciteBstWouldAddEndPuncttrue
\mciteSetBstMidEndSepPunct{\mcitedefaultmidpunct}
{\mcitedefaultendpunct}{\mcitedefaultseppunct}\relax
\EndOfBibitem
\bibitem[Fichou(1998)]{Fichou1999}
Fichou,~D. In \emph{{Handbook of Oligo- and Polythiophenes}}; Fichou,~D., Ed.;
  Wiley: Weinheim, Germany, 1998\relax
\mciteBstWouldAddEndPuncttrue
\mciteSetBstMidEndSepPunct{\mcitedefaultmidpunct}
{\mcitedefaultendpunct}{\mcitedefaultseppunct}\relax
\EndOfBibitem
\bibitem[Perepichka and Perepichka(2009)Perepichka, and
  Perepichka]{Perepichka2009}
Perepichka,~I.~F.; Perepichka,~D.~F. In \emph{{Handbook of Thiophene-Based
  Materials: Applications in Organic Electronics and Photonics}};
  Perepichka,~I.~F., Perepichka,~D.~F., Eds.; John Wiley {\&} Sons, Ltd:
  Chichester, UK, 2009\relax
\mciteBstWouldAddEndPuncttrue
\mciteSetBstMidEndSepPunct{\mcitedefaultmidpunct}
{\mcitedefaultendpunct}{\mcitedefaultseppunct}\relax
\EndOfBibitem
\bibitem[Fichou \latin{et~al.}(1992)Fichou, Horowitz, Xu, and
  Garnier]{Fichou1992}
Fichou,~D.; Horowitz,~G.; Xu,~B.; Garnier,~F. {Low temperature optical
  absorption of polycrystalline thin films of $\alpha$-quaterthiophene,
  $\alpha$-sexithiophene and $\alpha$-octithiophene, three model oligomers of
  polythiophene}. \emph{Synthetic Metals} \textbf{1992}, \emph{48},
  167--179\relax
\mciteBstWouldAddEndPuncttrue
\mciteSetBstMidEndSepPunct{\mcitedefaultmidpunct}
{\mcitedefaultendpunct}{\mcitedefaultseppunct}\relax
\EndOfBibitem
\bibitem[de~Melo \latin{et~al.}(1999)de~Melo, Silva, Arnaut, and
  Becker]{DeMelo1999}
de~Melo,~J.~S.; Silva,~L.~M.; Arnaut,~L.~G.; Becker,~R.~S. {Singlet and triplet
  energies of $\alpha$-oligothiophenes: A spectroscopic, theoretical, and
  photoacoustic study: Extrapolation to polythiophene}. \emph{The Journal of
  Chemical Physics} \textbf{1999}, \emph{111}, 5427--5433\relax
\mciteBstWouldAddEndPuncttrue
\mciteSetBstMidEndSepPunct{\mcitedefaultmidpunct}
{\mcitedefaultendpunct}{\mcitedefaultseppunct}\relax
\EndOfBibitem
\bibitem[Fabiano \latin{et~al.}(2005)Fabiano, Sala, Cingolani, Weimer, and
  G{\"{o}}rling]{Fabiano2005}
Fabiano,~E.; Sala,~F.~D.; Cingolani,~R.; Weimer,~M.; G{\"{o}}rling,~A.
  {Theoretical Study of Singlet and Triplet Excitation Energies in
  Oligothiophenes}. \emph{The Journal of Physical Chemistry A} \textbf{2005},
  \emph{109}, 3078--3085\relax
\mciteBstWouldAddEndPuncttrue
\mciteSetBstMidEndSepPunct{\mcitedefaultmidpunct}
{\mcitedefaultendpunct}{\mcitedefaultseppunct}\relax
\EndOfBibitem
\bibitem[Banks \latin{et~al.}(2005)Banks, Beard, Cao, Cho, Damm, Farid, Felts,
  Halgren, Mainz, Maple, Murphy, Philipp, Repasky, Zhang, Berne, Friesner,
  Gallicchio, and Levy]{Banks2005}
Banks,~J.~L. \latin{et~al.}  {Integrated Modeling Program, Applied Chemical
  Theory (IMPACT)}. \emph{Journal of Computational Chemistry} \textbf{2005},
  \emph{26}, 1752--1780\relax
\mciteBstWouldAddEndPuncttrue
\mciteSetBstMidEndSepPunct{\mcitedefaultmidpunct}
{\mcitedefaultendpunct}{\mcitedefaultseppunct}\relax
\EndOfBibitem
\bibitem[Fave(1992)]{Fave1992}
Fave,~J.-L. {Excitons in Chains of Thiophene Rings}. Electronic Properties of
  Polymers. Berlin, Heidelberg, 1992; pp 60--62\relax
\mciteBstWouldAddEndPuncttrue
\mciteSetBstMidEndSepPunct{\mcitedefaultmidpunct}
{\mcitedefaultendpunct}{\mcitedefaultseppunct}\relax
\EndOfBibitem
\bibitem[Pan \latin{et~al.}(2002)Pan, Chua, and Huang]{Pan2002}
Pan,~J.-F.; Chua,~S.-J.; Huang,~W. {Conformational analysis (ab initio
  HF/3-21G*) and optical properties of poly(thiophene-phenylene-thiophene)
  (PTPT)}. \emph{Chemical Physics Letters} \textbf{2002}, \emph{363},
  18--24\relax
\mciteBstWouldAddEndPuncttrue
\mciteSetBstMidEndSepPunct{\mcitedefaultmidpunct}
{\mcitedefaultendpunct}{\mcitedefaultseppunct}\relax
\EndOfBibitem
\bibitem[Westenhoff \latin{et~al.}(2006)Westenhoff, Beenken, Yartsev, and
  Greenham]{Westenhoff2006a}
Westenhoff,~S.; Beenken,~W. J.~D.; Yartsev,~A.; Greenham,~N.~C. {Conformational
  disorder of conjugated polymers}. \emph{The Journal of Chemical Physics}
  \textbf{2006}, \emph{125}, 154903\relax
\mciteBstWouldAddEndPuncttrue
\mciteSetBstMidEndSepPunct{\mcitedefaultmidpunct}
{\mcitedefaultendpunct}{\mcitedefaultseppunct}\relax
\EndOfBibitem
\bibitem[Darling(2008)]{Darling2008}
Darling,~S.~B. {Isolating the Effect of Torsional Defects on Mobility and Band
  Gap in Conjugated Polymers}. \emph{The Journal of Physical Chemistry B}
  \textbf{2008}, \emph{112}, 8891--8895\relax
\mciteBstWouldAddEndPuncttrue
\mciteSetBstMidEndSepPunct{\mcitedefaultmidpunct}
{\mcitedefaultendpunct}{\mcitedefaultseppunct}\relax
\EndOfBibitem
\bibitem[Dubay \latin{et~al.}(2012)Dubay, Hall, Hughes, Wu, Reichman, and
  Friesner]{Dubay2012}
Dubay,~K.~H.; Hall,~M.~L.; Hughes,~T.~F.; Wu,~C.; Reichman,~D.~R.;
  Friesner,~R.~a. {Accurate force field development for modeling conjugated
  polymers}. \emph{Journal of Chemical Theory and Computation} \textbf{2012},
  \emph{8}, 4556--4569\relax
\mciteBstWouldAddEndPuncttrue
\mciteSetBstMidEndSepPunct{\mcitedefaultmidpunct}
{\mcitedefaultendpunct}{\mcitedefaultseppunct}\relax
\EndOfBibitem
\bibitem[Bowers \latin{et~al.}(2006)Bowers, Sacerdoti, Salmon, Shan, Shaw,
  Chow, Xu, Dror, Eastwood, Gregersen, Klepeis, Kolossvary, and
  Moraes]{Bowers2006}
Bowers,~K.~J.; Sacerdoti,~F.~D.; Salmon,~J.~K.; Shan,~Y.; Shaw,~D.~E.;
  Chow,~E.; Xu,~H.; Dror,~R.~O.; Eastwood,~M.~P.; Gregersen,~B.~A.;
  Klepeis,~J.~L.; Kolossvary,~I.; Moraes,~M.~A. {Scalable algorithms for
  molecular dynamics simulations on commodity clusters}. Proceedings of the
  2006 ACM/IEEE conference on Supercomputing - SC '06. New York, New York, USA,
  2006; p~84\relax
\mciteBstWouldAddEndPuncttrue
\mciteSetBstMidEndSepPunct{\mcitedefaultmidpunct}
{\mcitedefaultendpunct}{\mcitedefaultseppunct}\relax
\EndOfBibitem
\bibitem[Nos{\'{e}}(1984)]{nose84b}
Nos{\'{e}},~S. {A molecular dynamics method for simulations in the canonical
  ensemble}. \emph{Molecular Physics} \textbf{1984}, \emph{52}, 255--268\relax
\mciteBstWouldAddEndPuncttrue
\mciteSetBstMidEndSepPunct{\mcitedefaultmidpunct}
{\mcitedefaultendpunct}{\mcitedefaultseppunct}\relax
\EndOfBibitem
\bibitem[Hoover(1985)]{hoover85}
Hoover,~W.~G. {Canonical dynamics: Equilibrium phase-space distributions}.
  \emph{Physical Review A} \textbf{1985}, \emph{31}, 1695--1697\relax
\mciteBstWouldAddEndPuncttrue
\mciteSetBstMidEndSepPunct{\mcitedefaultmidpunct}
{\mcitedefaultendpunct}{\mcitedefaultseppunct}\relax
\EndOfBibitem
\bibitem[Darden \latin{et~al.}(1993)Darden, York, and Pedersen]{darden93}
Darden,~T.; York,~D.; Pedersen,~L. {Particle mesh Ewald: An N.log(N) method for
  Ewald sums in large systems}. \emph{The Journal of Chemical Physics}
  \textbf{1993}, \emph{98}, 10089--10092\relax
\mciteBstWouldAddEndPuncttrue
\mciteSetBstMidEndSepPunct{\mcitedefaultmidpunct}
{\mcitedefaultendpunct}{\mcitedefaultseppunct}\relax
\EndOfBibitem
\bibitem[Essmann \latin{et~al.}(1995)Essmann, Perera, Berkowitz, Darden, Lee,
  and Pedersen]{essmann95}
Essmann,~U.; Perera,~L.; Berkowitz,~M.~L.; Darden,~T.; Lee,~H.; Pedersen,~L.~G.
  {A smooth particle mesh Ewald method}. \emph{The Journal of Chemical Physics}
  \textbf{1995}, \emph{103}, 8577--8593\relax
\mciteBstWouldAddEndPuncttrue
\mciteSetBstMidEndSepPunct{\mcitedefaultmidpunct}
{\mcitedefaultendpunct}{\mcitedefaultseppunct}\relax
\EndOfBibitem
\bibitem[Sun and Autschbach(2014)Sun, and Autschbach]{Sun2014}
Sun,~H.; Autschbach,~J. {Electronic Energy Gaps for $\pi$-Conjugated Oligomers
  and Polymers Calculated with Density Functional Theory}. \emph{Journal of
  Chemical Theory and Computation} \textbf{2014}, \emph{10}, 1035--1047\relax
\mciteBstWouldAddEndPuncttrue
\mciteSetBstMidEndSepPunct{\mcitedefaultmidpunct}
{\mcitedefaultendpunct}{\mcitedefaultseppunct}\relax
\EndOfBibitem
\bibitem[Sun \latin{et~al.}(2018)Sun, Berkelbach, Blunt, Booth, Guo, Li, Liu,
  McClain, Sayfutyarova, Sharma, Wouters, and Chan]{Sun2018}
Sun,~Q.; Berkelbach,~T.~C.; Blunt,~N.~S.; Booth,~G.~H.; Guo,~S.; Li,~Z.;
  Liu,~J.; McClain,~J.~D.; Sayfutyarova,~E.~R.; Sharma,~S.; Wouters,~S.;
  Chan,~G. K.-L. {PySCF: the Python-based simulations of chemistry framework}.
  \emph{Wiley Interdisciplinary Reviews: Computational Molecular Science}
  \textbf{2018}, \emph{8}, e1340\relax
\mciteBstWouldAddEndPuncttrue
\mciteSetBstMidEndSepPunct{\mcitedefaultmidpunct}
{\mcitedefaultendpunct}{\mcitedefaultseppunct}\relax
\EndOfBibitem
\bibitem[Sun(2015)]{Sun2015}
Sun,~Q. {Libcint: An efficient general integral library for Gaussian basis
  functions}. \emph{Journal of Computational Chemistry} \textbf{2015},
  \emph{36}, 1664--1671\relax
\mciteBstWouldAddEndPuncttrue
\mciteSetBstMidEndSepPunct{\mcitedefaultmidpunct}
{\mcitedefaultendpunct}{\mcitedefaultseppunct}\relax
\EndOfBibitem
\bibitem[Hansen \latin{et~al.}(2015)Hansen, Biegler, Ramakrishnan, Pronobis,
  von Lilienfeld, M{\"{u}}ller, and Tkatchenko]{Hansen2015}
Hansen,~K.; Biegler,~F.; Ramakrishnan,~R.; Pronobis,~W.; von Lilienfeld,~O.~A.;
  M{\"{u}}ller,~K.-R.; Tkatchenko,~A. {Machine Learning Predictions of
  Molecular Properties: Accurate Many-Body Potentials and Nonlocality in
  Chemical Space}. \emph{The Journal of Physical Chemistry Letters}
  \textbf{2015}, \emph{6}, 2326--2331\relax
\mciteBstWouldAddEndPuncttrue
\mciteSetBstMidEndSepPunct{\mcitedefaultmidpunct}
{\mcitedefaultendpunct}{\mcitedefaultseppunct}\relax
\EndOfBibitem
\bibitem[Bart{\'{o}}k \latin{et~al.}(2013)Bart{\'{o}}k, Kondor, and
  Cs{\'{a}}nyi]{Bartok2013}
Bart{\'{o}}k,~A.~P.; Kondor,~R.; Cs{\'{a}}nyi,~G. {On representing chemical
  environments}. \emph{Physical Review B} \textbf{2013}, \emph{87},
  184115\relax
\mciteBstWouldAddEndPuncttrue
\mciteSetBstMidEndSepPunct{\mcitedefaultmidpunct}
{\mcitedefaultendpunct}{\mcitedefaultseppunct}\relax
\EndOfBibitem
\bibitem[Behler(2011)]{Behler2011}
Behler,~J. {Atom-centered symmetry functions for constructing high-dimensional
  neural network potentials}. \emph{The Journal of Chemical Physics}
  \textbf{2011}, \emph{134}, 074106\relax
\mciteBstWouldAddEndPuncttrue
\mciteSetBstMidEndSepPunct{\mcitedefaultmidpunct}
{\mcitedefaultendpunct}{\mcitedefaultseppunct}\relax
\EndOfBibitem
\bibitem[Sch{\"{u}}tt \latin{et~al.}(2017)Sch{\"{u}}tt, Kindermans, Felix,
  Chmiela, Tkatchenko, and M{\"{u}}ller]{Schuett2017a}
Sch{\"{u}}tt,~K.; Kindermans,~P.-J.; Felix,~H. E.~S.; Chmiela,~S.;
  Tkatchenko,~A.; M{\"{u}}ller,~K.-R. {Schnet: A continuous-filter
  convolutional neural network for modeling quantum interactions}. Advances in
  Neural Information Processing Systems. 2017; pp 991--1001\relax
\mciteBstWouldAddEndPuncttrue
\mciteSetBstMidEndSepPunct{\mcitedefaultmidpunct}
{\mcitedefaultendpunct}{\mcitedefaultseppunct}\relax
\EndOfBibitem
\bibitem[alc()]{alchemy}
{Source Codes}.
  \url{https://github.com/tencent-alchemy/Alchemy/tree/master/dgl}\relax
\mciteBstWouldAddEndPuncttrue
\mciteSetBstMidEndSepPunct{\mcitedefaultmidpunct}
{\mcitedefaultendpunct}{\mcitedefaultseppunct}\relax
\EndOfBibitem
\bibitem[Ramsundar \latin{et~al.}(2019)Ramsundar, Eastman, Walters, Pande,
  Leswing, and Wu]{Ramsundar2019}
Ramsundar,~B.; Eastman,~P.; Walters,~P.; Pande,~V.; Leswing,~K.; Wu,~Z.
  \emph{{Deep Learning for the Life Sciences}}; O'Reilly Media, 2019\relax
\mciteBstWouldAddEndPuncttrue
\mciteSetBstMidEndSepPunct{\mcitedefaultmidpunct}
{\mcitedefaultendpunct}{\mcitedefaultseppunct}\relax
\EndOfBibitem
\bibitem[Dreuw and Head-Gordon(2005)Dreuw, and Head-Gordon]{Dreuw2005}
Dreuw,~A.; Head-Gordon,~M. {Single-Reference ab Initio Methods for the
  Calculation of Excited States of Large Molecules}. \emph{Chemical Reviews}
  \textbf{2005}, \emph{105}, 4009--4037\relax
\mciteBstWouldAddEndPuncttrue
\mciteSetBstMidEndSepPunct{\mcitedefaultmidpunct}
{\mcitedefaultendpunct}{\mcitedefaultseppunct}\relax
\EndOfBibitem
\bibitem[Reed \latin{et~al.}(1985)Reed, Weinstock, and Weinhold]{Reed1985}
Reed,~A.~E.; Weinstock,~R.~B.; Weinhold,~F. {Natural population analysis}.
  \emph{The Journal of Chemical Physics} \textbf{1985}, \emph{83},
  735--746\relax
\mciteBstWouldAddEndPuncttrue
\mciteSetBstMidEndSepPunct{\mcitedefaultmidpunct}
{\mcitedefaultendpunct}{\mcitedefaultseppunct}\relax
\EndOfBibitem
\bibitem[Segatta \latin{et~al.}(2019)Segatta, Cupellini, Garavelli, and
  Mennucci]{Segatta2019}
Segatta,~F.; Cupellini,~L.; Garavelli,~M.; Mennucci,~B. {Quantum Chemical
  Modeling of the Photoinduced Activity of Multichromophoric Biosystems}.
  \emph{Chemical Reviews} \textbf{2019}, \emph{119}, 9361--9380\relax
\mciteBstWouldAddEndPuncttrue
\mciteSetBstMidEndSepPunct{\mcitedefaultmidpunct}
{\mcitedefaultendpunct}{\mcitedefaultseppunct}\relax
\EndOfBibitem
\bibitem[Loco and Cupellini(2019)Loco, and Cupellini]{Loco2019}
Loco,~D.; Cupellini,~L. {Modeling the absorption lineshape of embedded systems
  from molecular dynamics: A tutorial review}. \emph{International Journal of
  Quantum Chemistry} \textbf{2019}, \emph{119}, e25726\relax
\mciteBstWouldAddEndPuncttrue
\mciteSetBstMidEndSepPunct{\mcitedefaultmidpunct}
{\mcitedefaultendpunct}{\mcitedefaultseppunct}\relax
\EndOfBibitem
\bibitem[Zuehlsdorff \latin{et~al.}(2019)Zuehlsdorff, Montoya-Castillo, Napoli,
  Markland, and Isborn]{Zuehlsdorff2019a}
Zuehlsdorff,~T.~J.; Montoya-Castillo,~A.; Napoli,~J.~A.; Markland,~T.~E.;
  Isborn,~C.~M. {Optical spectra in the condensed phase: Capturing anharmonic
  and vibronic features using dynamic and static approaches}. \emph{The Journal
  of Chemical Physics} \textbf{2019}, \emph{151}, 074111\relax
\mciteBstWouldAddEndPuncttrue
\mciteSetBstMidEndSepPunct{\mcitedefaultmidpunct}
{\mcitedefaultendpunct}{\mcitedefaultseppunct}\relax
\EndOfBibitem
\bibitem[Zuehlsdorff and Isborn(2019)Zuehlsdorff, and Isborn]{Zuehlsdorff2019}
Zuehlsdorff,~T.~J.; Isborn,~C.~M. {Modeling absorption spectra of molecules in
  solution}. \emph{International Journal of Quantum Chemistry} \textbf{2019},
  \emph{119}, e25719\relax
\mciteBstWouldAddEndPuncttrue
\mciteSetBstMidEndSepPunct{\mcitedefaultmidpunct}
{\mcitedefaultendpunct}{\mcitedefaultseppunct}\relax
\EndOfBibitem
\bibitem[Zuehlsdorff and Isborn(2018)Zuehlsdorff, and Isborn]{Zuehlsdorff2018}
Zuehlsdorff,~T.~J.; Isborn,~C.~M. {Combining the ensemble and Franck-Condon
  approaches for calculating spectral shapes of molecules in solution}.
  \emph{The Journal of Chemical Physics} \textbf{2018}, \emph{148},
  024110\relax
\mciteBstWouldAddEndPuncttrue
\mciteSetBstMidEndSepPunct{\mcitedefaultmidpunct}
{\mcitedefaultendpunct}{\mcitedefaultseppunct}\relax
\EndOfBibitem
\bibitem[Shi and Willard(2018)Shi, and Willard]{Shi2018}
Shi,~L.; Willard,~A.~P. {Modeling the effects of molecular disorder on the
  properties of Frenkel excitons in organic molecular semiconductors}.
  \emph{The Journal of Chemical Physics} \textbf{2018}, \emph{149},
  094110\relax
\mciteBstWouldAddEndPuncttrue
\mciteSetBstMidEndSepPunct{\mcitedefaultmidpunct}
{\mcitedefaultendpunct}{\mcitedefaultseppunct}\relax
\EndOfBibitem
\bibitem[McQuarrie(1976)]{mcquarrie76}
McQuarrie,~D.~A. \emph{{Statistical Mechanics}}; Harper and Row: New York,
  1976\relax
\mciteBstWouldAddEndPuncttrue
\mciteSetBstMidEndSepPunct{\mcitedefaultmidpunct}
{\mcitedefaultendpunct}{\mcitedefaultseppunct}\relax
\EndOfBibitem
\bibitem[Mukamel(1995)]{mukamel95}
Mukamel,~S. \emph{{Principles of Nonlinear Optical Spectroscopy}}; Oxford: New
  York, 1995\relax
\mciteBstWouldAddEndPuncttrue
\mciteSetBstMidEndSepPunct{\mcitedefaultmidpunct}
{\mcitedefaultendpunct}{\mcitedefaultseppunct}\relax
\EndOfBibitem
\bibitem[Becker \latin{et~al.}(1996)Becker, {Seixas de Melo}, Ma{\c{c}}anita,
  and Elisei]{Becker1996}
Becker,~R.~S.; {Seixas de Melo},~J.; Ma{\c{c}}anita,~A.~L.; Elisei,~F.
  {Comprehensive Evaluation of the Absorption, Photophysical, Energy Transfer,
  Structural, and Theoretical Properties of $\alpha$-Oligothiophenes with One
  to Seven Rings}. \emph{The Journal of Physical Chemistry} \textbf{1996},
  \emph{100}, 18683--18695\relax
\mciteBstWouldAddEndPuncttrue
\mciteSetBstMidEndSepPunct{\mcitedefaultmidpunct}
{\mcitedefaultendpunct}{\mcitedefaultseppunct}\relax
\EndOfBibitem
\bibitem[Egelhaaf \latin{et~al.}(1998)Egelhaaf, Oelkrug, Gebauer, Sokolowski,
  Umbach, Fischer, and B{\"{a}}uerle]{Egelhaaf1998}
Egelhaaf,~H.-J.; Oelkrug,~D.; Gebauer,~W.; Sokolowski,~M.; Umbach,~E.;
  Fischer,~T.; B{\"{a}}uerle,~P. {Photophysical properties of $\beta$-alkylated
  quater-, octa-, dodeca- and hexadecatiophenes}. \emph{Optical Materials}
  \textbf{1998}, \emph{9}, 59--64\relax
\mciteBstWouldAddEndPuncttrue
\mciteSetBstMidEndSepPunct{\mcitedefaultmidpunct}
{\mcitedefaultendpunct}{\mcitedefaultseppunct}\relax
\EndOfBibitem
\bibitem[Taliani and Gebauer(2007)Taliani, and Gebauer]{Taliani}
Taliani,~C.; Gebauer,~W. \emph{Handbook of Oligo- and Polythiophenes};
  Wiley-VCH Verlag GmbH: Weinheim, Germany, 2007; pp 361--404\relax
\mciteBstWouldAddEndPuncttrue
\mciteSetBstMidEndSepPunct{\mcitedefaultmidpunct}
{\mcitedefaultendpunct}{\mcitedefaultseppunct}\relax
\EndOfBibitem
\bibitem[Grebner \latin{et~al.}(1995)Grebner, Helbig, and Rentsch]{Grebner1995}
Grebner,~D.; Helbig,~M.; Rentsch,~S. {Size-dependent properties of
  oligothiophenes by picosecond time-resolved spectroscopy}. \emph{Journal of
  Physical Chemistry} \textbf{1995}, \emph{99}, 16991--16998\relax
\mciteBstWouldAddEndPuncttrue
\mciteSetBstMidEndSepPunct{\mcitedefaultmidpunct}
{\mcitedefaultendpunct}{\mcitedefaultseppunct}\relax
\EndOfBibitem
\bibitem[Lap \latin{et~al.}(1997)Lap, Grebner, and Rentsch]{Lap1997}
Lap,~D.~V.; Grebner,~D.; Rentsch,~S. {Femtosecond Time-Resolved Spectroscopic
  Studies on Thiophene Oligomers}. \emph{The Journal of Physical Chemistry A}
  \textbf{1997}, \emph{101}, 107--112\relax
\mciteBstWouldAddEndPuncttrue
\mciteSetBstMidEndSepPunct{\mcitedefaultmidpunct}
{\mcitedefaultendpunct}{\mcitedefaultseppunct}\relax
\EndOfBibitem
\bibitem[Zhao \latin{et~al.}(1988)Zhao, Singh, and Prasad]{Zhao1988}
Zhao,~M.; Singh,~B.~P.; Prasad,~P.~N. {A systematic study of polarizability and
  microscopic third-order optical nonlinearity in thiophene oligomers}.
  \emph{The Journal of Chemical Physics} \textbf{1988}, \emph{89},
  5535--5541\relax
\mciteBstWouldAddEndPuncttrue
\mciteSetBstMidEndSepPunct{\mcitedefaultmidpunct}
{\mcitedefaultendpunct}{\mcitedefaultseppunct}\relax
\EndOfBibitem
\bibitem[Sease and Zechmeister(1947)Sease, and Zechmeister]{Sease1947}
Sease,~J.~W.; Zechmeister,~L. {Chromatographic and Spectral Characteristics of
  Some Polythienyls}. \emph{Journal of the American Chemical Society}
  \textbf{1947}, \emph{69}, 270--273\relax
\mciteBstWouldAddEndPuncttrue
\mciteSetBstMidEndSepPunct{\mcitedefaultmidpunct}
{\mcitedefaultendpunct}{\mcitedefaultseppunct}\relax
\EndOfBibitem
\bibitem[Salzner(2007)]{Salzner2007}
Salzner,~U. {Theoretical Investigation of Excited States of Oligothiophenes and
  of Their Monocations}. \emph{Journal of Chemical Theory and Computation}
  \textbf{2007}, \emph{3}, 1143--1157\relax
\mciteBstWouldAddEndPuncttrue
\mciteSetBstMidEndSepPunct{\mcitedefaultmidpunct}
{\mcitedefaultendpunct}{\mcitedefaultseppunct}\relax
\EndOfBibitem
\bibitem[Alem{\'{a}}n \latin{et~al.}(2011)Alem{\'{a}}n, Torras, and
  Casanovas]{Aleman2011}
Alem{\'{a}}n,~C.; Torras,~J.; Casanovas,~J. {Influence of polarity of the
  medium in the saturation of the electronic properties for $\pi$-conjugated
  oligothiophenes}. \emph{Chemical Physics Letters} \textbf{2011}, \emph{511},
  283--287\relax
\mciteBstWouldAddEndPuncttrue
\mciteSetBstMidEndSepPunct{\mcitedefaultmidpunct}
{\mcitedefaultendpunct}{\mcitedefaultseppunct}\relax
\EndOfBibitem
\bibitem[Lee \latin{et~al.}(2019)Lee, Shi, and Willard]{Lee2019}
Lee,~C.~K.; Shi,~L.; Willard,~A.~P. {Modeling the Influence of Correlated
  Molecular Disorder on the Dynamics of Excitons in Organic Molecular
  Semiconductors}. \emph{The Journal of Physical Chemistry C} \textbf{2019},
  \emph{123}, 306--314\relax
\mciteBstWouldAddEndPuncttrue
\mciteSetBstMidEndSepPunct{\mcitedefaultmidpunct}
{\mcitedefaultendpunct}{\mcitedefaultseppunct}\relax
\EndOfBibitem
\bibitem[Wu \latin{et~al.}(2019)Wu, Kondo, Kakimoto, Yang, Yamada, Kuwajima,
  Lambard, Hongo, Xu, Shiomi, Schick, Morikawa, and Yoshida]{Wu2019}
Wu,~S.; Kondo,~Y.; Kakimoto,~M.-a.; Yang,~B.; Yamada,~H.; Kuwajima,~I.;
  Lambard,~G.; Hongo,~K.; Xu,~Y.; Shiomi,~J.; Schick,~C.; Morikawa,~J.;
  Yoshida,~R. {Machine-learning-assisted discovery of polymers with high
  thermal conductivity using a molecular design algorithm}. \emph{npj
  Computational Materials} \textbf{2019}, \emph{5}, 66\relax
\mciteBstWouldAddEndPuncttrue
\mciteSetBstMidEndSepPunct{\mcitedefaultmidpunct}
{\mcitedefaultendpunct}{\mcitedefaultseppunct}\relax
\EndOfBibitem
\bibitem[Smith \latin{et~al.}(2019)Smith, Nebgen, Zubatyuk, Lubbers, Devereux,
  Barros, Tretiak, Isayev, and Roitberg]{Smith2019}
Smith,~J.~S.; Nebgen,~B.~T.; Zubatyuk,~R.; Lubbers,~N.; Devereux,~C.;
  Barros,~K.; Tretiak,~S.; Isayev,~O.; Roitberg,~A.~E. {Approaching coupled
  cluster accuracy with a general-purpose neural network potential through
  transfer learning}. \emph{Nature Communications} \textbf{2019}, \emph{10},
  2903\relax
\mciteBstWouldAddEndPuncttrue
\mciteSetBstMidEndSepPunct{\mcitedefaultmidpunct}
{\mcitedefaultendpunct}{\mcitedefaultseppunct}\relax
\EndOfBibitem
\bibitem[Yamada \latin{et~al.}(2019)Yamada, Liu, Wu, Koyama, Ju, Shiomi,
  Morikawa, and Yoshida]{Yamada2019}
Yamada,~H.; Liu,~C.; Wu,~S.; Koyama,~Y.; Ju,~S.; Shiomi,~J.; Morikawa,~J.;
  Yoshida,~R. {Predicting Materials Properties with Little Data Using Shotgun
  Transfer Learning}. \emph{ACS Central Science} \textbf{2019}, \emph{5},
  1717--1730\relax
\mciteBstWouldAddEndPuncttrue
\mciteSetBstMidEndSepPunct{\mcitedefaultmidpunct}
{\mcitedefaultendpunct}{\mcitedefaultseppunct}\relax
\EndOfBibitem
\end{mcitethebibliography}


\begin{thebibliography}{10}
\expandafter\ifx\csname url\endcsname\relax
  \def\url#1{\texttt{#1}}\fi
\expandafter\ifx\csname urlprefix\endcsname\relax\def\urlprefix{URL }\fi
\providecommand{\bibinfo}[2]{#2}
\providecommand{\eprint}[2][]{\url{#2}}

\bibitem{kingma2015adam}
\bibinfo{author}{Kingma, D.~P.} \& \bibinfo{author}{Ba, J.}
\newblock \bibinfo{title}{{Adam: A method for stochastic optimization}}.
\newblock In \emph{\bibinfo{booktitle}{International Conference on Learning
  Representations (ICLR)}} (\bibinfo{year}{2015}).

\bibitem{paszke2017pytorch}
\bibinfo{author}{Adam, P.} \emph{et~al.}
\newblock \bibinfo{title}{{Automatic differentiation in pytorch}}.
\newblock In \emph{\bibinfo{booktitle}{Proceedings of Neural Information
  Processing Systems}} (\bibinfo{year}{2017}).

\bibitem{wang2019dgl}
\bibinfo{author}{Wang, M.} \emph{et~al.}
\newblock \bibinfo{title}{{Deep Graph Library: Towards Efficient and Scalable
  Deep Learning on Graphs}}.
\newblock \emph{\bibinfo{journal}{ICLR Workshop on Representation Learning on
  Graphs and Manifolds}}  (\bibinfo{year}{2019}).

\bibitem{alchemy}
\bibinfo{title}{{Source Codes}}.
\newblock
  \urlprefix\url{https://github.com/tencent-alchemy/Alchemy/tree/master/dgl}.

\bibitem{Ramsundar2019}
\bibinfo{author}{Ramsundar, B.} \emph{et~al.}
\newblock \emph{\bibinfo{title}{{Deep Learning for the Life Sciences}}}
  (\bibinfo{publisher}{O'Reilly Media}, \bibinfo{year}{2019}).

\bibitem{Salzner2011}
\bibinfo{author}{Salzner, U.} \& \bibinfo{author}{Aydin, A.}
\newblock \bibinfo{title}{{Improved Prediction of Properties of
  $\pi$-Conjugated Oligomers with Range-Separated Hybrid Density Functionals}}.
\newblock \emph{\bibinfo{journal}{Journal of Chemical Theory and Computation}}
  \textbf{\bibinfo{volume}{7}}, \bibinfo{pages}{2568--2583}
  (\bibinfo{year}{2011}).

\bibitem{Salzner2014}
\bibinfo{author}{Salzner, U.}
\newblock \bibinfo{title}{{Electronic structure of conducting organic polymers:
  insights from time-dependent density functional theory}}.
\newblock \emph{\bibinfo{journal}{Wiley Interdisciplinary Reviews:
  Computational Molecular Science}} \textbf{\bibinfo{volume}{4}},
  \bibinfo{pages}{601--622} (\bibinfo{year}{2014}).

\bibitem{Fabiano2005}
\bibinfo{author}{Fabiano, E.}, \bibinfo{author}{Sala, F.~D.},
  \bibinfo{author}{Cingolani, R.}, \bibinfo{author}{Weimer, M.} \&
  \bibinfo{author}{G{\"{o}}rling, A.}
\newblock \bibinfo{title}{{Theoretical Study of Singlet and Triplet Excitation
  Energies in Oligothiophenes}}.
\newblock \emph{\bibinfo{journal}{The Journal of Physical Chemistry A}}
  \textbf{\bibinfo{volume}{109}}, \bibinfo{pages}{3078--3085}
  (\bibinfo{year}{2005}).

\bibitem{Jacquemin2015}
\bibinfo{author}{Jacquemin, D.}, \bibinfo{author}{Duchemin, I.} \&
  \bibinfo{author}{Blase, X.}
\newblock \bibinfo{title}{{0-0 Energies Using Hybrid Schemes: Benchmarks of
  TD-DFT, CIS(D), ADC(2), CC2, and BSE/ GW formalisms for 80 Real-Life
  Compounds}}.
\newblock \emph{\bibinfo{journal}{Journal of Chemical Theory and Computation}}
  \textbf{\bibinfo{volume}{11}}, \bibinfo{pages}{5340--5359}
  (\bibinfo{year}{2015}).

\bibitem{Dreuw2005}
\bibinfo{author}{Dreuw, A.} \& \bibinfo{author}{Head-Gordon, M.}
\newblock \bibinfo{title}{{Single-Reference ab Initio Methods for the
  Calculation of Excited States of Large Molecules}}.
\newblock \emph{\bibinfo{journal}{Chemical Reviews}}
  \textbf{\bibinfo{volume}{105}}, \bibinfo{pages}{4009--4037}
  (\bibinfo{year}{2005}).

\bibitem{Sun2018}
\bibinfo{author}{Sun, Q.} \emph{et~al.}
\newblock \bibinfo{title}{{PySCF: the Python-based simulations of chemistry
  framework}}.
\newblock \emph{\bibinfo{journal}{Wiley Interdisciplinary Reviews:
  Computational Molecular Science}} \textbf{\bibinfo{volume}{8}},
  \bibinfo{pages}{e1340} (\bibinfo{year}{2018}).

\bibitem{Sun2015}
\bibinfo{author}{Sun, Q.}
\newblock \bibinfo{title}{{Libcint: An efficient general integral library for
  Gaussian basis functions}}.
\newblock \emph{\bibinfo{journal}{Journal of Computational Chemistry}}
  \textbf{\bibinfo{volume}{36}}, \bibinfo{pages}{1664--1671}
  (\bibinfo{year}{2015}).

\bibitem{Bowers2006}
\bibinfo{author}{Bowers, K.~J.} \emph{et~al.}
\newblock \bibinfo{title}{{Scalable algorithms for molecular dynamics
  simulations on commodity clusters}}.
\newblock In \emph{\bibinfo{booktitle}{Proceedings of the 2006 ACM/IEEE
  conference on Supercomputing - SC '06}}, \bibinfo{pages}{84}
  (\bibinfo{publisher}{ACM Press}, \bibinfo{address}{New York, New York, USA},
  \bibinfo{year}{2006}).

\bibitem{Martyna1994}
\bibinfo{author}{Martyna, G.~J.}, \bibinfo{author}{Tobias, D.~J.} \&
  \bibinfo{author}{Klein, M.~L.}
\newblock \bibinfo{title}{{Constant pressure molecular dynamics algorithms}}.
\newblock \emph{\bibinfo{journal}{The Journal of Chemical Physics}}
  \textbf{\bibinfo{volume}{101}}, \bibinfo{pages}{4177} (\bibinfo{year}{1994}).

\bibitem{Taliani}
\bibinfo{author}{Taliani, C.} \& \bibinfo{author}{Gebauer, W.}
\newblock \bibinfo{title}{{Electronic Excited States of Conjugated
  Oligothiophenes}}.
\newblock In \emph{\bibinfo{booktitle}{Handbook of Oligo- and Polythiophenes}},
  \bibinfo{pages}{361--404} (\bibinfo{publisher}{Wiley-VCH Verlag GmbH},
  \bibinfo{address}{Weinheim, Germany}, \bibinfo{year}{2007}).
\newblock \urlprefix\url{http://doi.wiley.com/10.1002/9783527611713.ch7}.

\bibitem{Grebner1995}
\bibinfo{author}{Grebner, D.}, \bibinfo{author}{Helbig, M.} \&
  \bibinfo{author}{Rentsch, S.}
\newblock \bibinfo{title}{{Size-dependent properties of oligothiophenes by
  picosecond time-resolved spectroscopy}}.
\newblock \emph{\bibinfo{journal}{Journal of Physical Chemistry}}
  \textbf{\bibinfo{volume}{99}}, \bibinfo{pages}{16991--16998}
  (\bibinfo{year}{1995}).

\bibitem{Lap1997}
\bibinfo{author}{Lap, D.~V.}, \bibinfo{author}{Grebner, D.} \&
  \bibinfo{author}{Rentsch, S.}
\newblock \bibinfo{title}{{Femtosecond Time-Resolved Spectroscopic Studies on
  Thiophene Oligomers}}.
\newblock \emph{\bibinfo{journal}{The Journal of Physical Chemistry A}}
  \textbf{\bibinfo{volume}{101}}, \bibinfo{pages}{107--112}
  (\bibinfo{year}{1997}).

\bibitem{Zhao1988}
\bibinfo{author}{Zhao, M.}, \bibinfo{author}{Singh, B.~P.} \&
  \bibinfo{author}{Prasad, P.~N.}
\newblock \bibinfo{title}{{A systematic study of polarizability and microscopic
  third-order optical nonlinearity in thiophene oligomers}}.
\newblock \emph{\bibinfo{journal}{The Journal of Chemical Physics}}
  \textbf{\bibinfo{volume}{89}}, \bibinfo{pages}{5535--5541}
  (\bibinfo{year}{1988}).

\bibitem{Sease1947}
\bibinfo{author}{Sease, J.~W.} \& \bibinfo{author}{Zechmeister, L.}
\newblock \bibinfo{title}{{Chromatographic and Spectral Characteristics of Some
  Polythienyls}}.
\newblock \emph{\bibinfo{journal}{Journal of the American Chemical Society}}
  \textbf{\bibinfo{volume}{69}}, \bibinfo{pages}{270--273}
  (\bibinfo{year}{1947}).

\bibitem{Salzner2007}
\bibinfo{author}{Salzner, U.}
\newblock \bibinfo{title}{{Theoretical Investigation of Excited States of
  Oligothiophenes and of Their Monocations}}.
\newblock \emph{\bibinfo{journal}{Journal of Chemical Theory and Computation}}
  \textbf{\bibinfo{volume}{3}}, \bibinfo{pages}{1143--1157}
  (\bibinfo{year}{2007}).

\bibitem{Aleman2011}
\bibinfo{author}{Alem{\'{a}}n, C.}, \bibinfo{author}{Torras, J.} \&
  \bibinfo{author}{Casanovas, J.}
\newblock \bibinfo{title}{{Influence of polarity of the medium in the
  saturation of the electronic properties for $\pi$-conjugated
  oligothiophenes}}.
\newblock \emph{\bibinfo{journal}{Chemical Physics Letters}}
  \textbf{\bibinfo{volume}{511}}, \bibinfo{pages}{283--287}
  (\bibinfo{year}{2011}).

\end{thebibliography}

\end{document}


\newcommand{\wn}{cm$^{-1}$}
\newcommand{\td}{$\sim$}
\newcommand{\la}{\langle}
\newcommand{\ra}{\rangle}
\newcommand{\e}{\epsilon}
\newcommand{\w}{\omega}
\newcommand{\bracket}[1]{\left\langle #1 \right\rangle}
\newcommand{\degreec}{^{\circ}{\rm C}}
\newcommand{\be}{\begin{equation}}
\newcommand{\ee}{\end{equation}}
\newcommand{\ie}{{\it i.e.}}
\newcommand{\eg}{{\it e.g.}}
\newcommand{\etal}{{\it et al.}}
\newcommand{\bra}[1]{\left<#1\right|}
\newcommand{\ket}[1]{\left|#1\right>}
\newcommand{\ketbra}[2]{\ket{#1}\bra{#2}}

\title{{\huge Supporting Information}\\
\vskip 0.2in
Deep Learning for Optoelectronic Properties of Organic Semiconductors}
\author{Chengqiang Lu}
\affiliation{Anhui Province Key Lab of Big Data Analysis and Application, University of Science and Technology of China, Hefei, Anhui 230026, China}
\author{Qi Liu}
\email{qiliuql@ustc.edu.cn}
\affiliation{Anhui Province Key Lab of Big Data Analysis and Application, University of Science and Technology of China, Hefei, Anhui 230026, China}
\author{Qiming Sun}
\affiliation{Tencent America, Palo Alto, CA 94306 , United States}
\author{Chang-Yu Hsieh}
\affiliation{Tencent, Shenzhen, Guangdong 518057, China, }
\author{Shengyu Zhang}
\affiliation{Tencent, Shenzhen, Guangdong 518057, China, }
\author{Liang Shi}
\email{lshi4@ucmerced.edu}
\affiliation{Chemistry and Chemical Biology, University of California, Merced, California 95343, United States}
\author{Chee-Kong Lee}
\email{cheekonglee@tencent.com}
\affiliation{Tencent America, Palo Alto, CA 94306, United States}
\maketitle
\setstretch{1.25}

\section{Implementation Details of Deep Neural Networks}
In this section, we will provide additional implementation details of the deep neural networks (DNNs) used in the main text. As mentioned in the main text, we use the same hyperparameters as the original papers, unless otherwise stated. Particularly for SchNet, a cutoff of 5\AA\ is used for the predictions of all electronic properties except the transition dipole moments, for which we use a cutoff of 25 \AA. The dimension of embedding in SchNet is set to be 128, and we use three interaction blocks.

We train all DNN models with Adam optimizer\cite{kingma2015adam} using a learning rate of 0.0001 and a batch size of 64. For SchNet, the decay rate is set to 1 (i.e., no decay) since we found that adding decay does not lead to noticeable improvement. We train all the models for a maximum of 750 epochs and use the validation set for early-stopping. The DNNs are implemented using Pytorch \cite{paszke2017pytorch} and Deep Graph Library \cite{wang2019dgl}. The codes of our implementations of MGCN, SchNet and MPNN can be found on Github~\cite{alchemy}, and we use DeepChem for DTNN~\cite{Ramsundar2019}.  

\section{Benchmark of DFT For Oligothiophenes}

For organic semiconductors, range-separated hybrid density functionals are often required to reduce the self-interaction error, and Salzner and co-workers have conducted comprehensive benchmark studies on OTs~\cite{Salzner2011,Salzner2014}. They found that $\omega$B97XD and CAM-B3LYP show the best overall performance of all range-separated functionals they tested~\cite{Salzner2011}. To further benchmark our DFT method, we computed the lowest singlet excited-state energies of isolated OTs against the reported results from a correlated wavefunction method (CC2 method) in Ref. \onlinecite{Fabiano2005}. To make a fair comparison with the CC2 results, we optimized the geometries of OTs with planar constraints using B3LYP/cc-pVTZ following Ref. \onlinecite{Fabiano2005}. TDDFT calculations were then performed with the Tamm-Dancoff approximation (TDA) at the level of CAM-B3LYP/6-31+G(d) to obtain the lowest singlet excited state energies. All the calculations were performed using PySCF. As shown in Table \ref{tab:ot}, the predictions from CAM-B3LYP/6-31+G(d) are within 0.1 eV of those from CC2, whose error with respect to experimental estimate is shown to be around 0.15 eV in average for the lowest lying excited-state energies of medium to large molecules~\cite{Jacquemin2015}. Based on previous studies\cite{Salzner2011,Salzner2014,Jacquemin2015} and the benchmark results here, we estimate the average error of our TDDFT calculation in predicting the lowest lying excited-state energies of OTs to be around 0.2-0.3eV, within the range of the typical error of TDDFT (0.1-0.5eV)~\cite{Dreuw2005}. 


\renewcommand{\thetable}{S1}
\begin{table}[h!tbp]
\begin{tabular}{cccccc}
\hline
Method & 2T & 3T & 4T & 5T & 6T  \\
\hline
CC2\cite{Fabiano2005}         & 4.26 & 3.57 & 3.21 & 2.96 &  -    \\
CAM-B3LYP   & 4.34 & 3.63 & 3.24 & 3.00 & 2.84  \\
\hline
\end{tabular}
\caption{Calculated excitation energies in eV for the lowest singlet excited state of isolated OTs. The results from the second-order approximate coupled cluster singles and doubles (CC2) are taken from Ref. \onlinecite{Fabiano2005}. The geometries of OTs are optimized at the level of B3LYP/cc-pVTZ with the constraint of planarity following Ref. \onlinecite{Fabiano2005}. TDDFT calculations with the CAM-B3LYP functional are performed with the basis set 6-31+G(d) using PySCF~\cite{Sun2018,Sun2015}.}
\label{tab:ot}
\end{table}


\section{More Discussions on Absorption Spectral Simulations}

For the computation of the UV-Vis absorption spectrum of OT in dichloromethane, a single OT molecule was immersed in a simulation box of about 530 dichloromethane molecules, and classical MD simulation was performed in the NPT ensemble at 300 K and 1 atm using Desmond~\cite{Bowers2006}. Martyna-Tobias-Klein scheme~\cite{Martyna1994} was employed to maintain temperature and pressure with a coupling constant of 2.0 ps for both. The other simulation details are the same as those for isolated OTs described in Section 2 of the main text. Configurations of OTs were saved every 10 fs over a 10-ns trajectory for each OT, and then were fed into the SchNet models to compute the energies and transition dipole moments for the two lowest lying singlet states. Note that our SchNet models are trained against TDDFT results of isolated OTs. By using these SchNet models for OTs in solutions, we have neglected the effects of solvent on the excited-state properties of OTs, a reasonable approximation based on previous studies on OTs~\cite{Taliani,Grebner1995,Lap1997,Zhao1988,Sease1947,Salzner2007,Aleman2011}. Table \ref{tab:abs} show the absorption maxima of the experimental absorption spectra of OTs in organic solvents of varying polarity, and it is clear that the solvatochromism of OTs is fairly small. It has been also shown theoretically that the solvent effects on OTs are small~\cite{Salzner2007,Aleman2011}.

In the experimental spectra of OTs in dichloromethane, there are extra peaks or shoulders on the blue side of the main peaks, and we attribute them to higher excited states. To verify this, we computed the first five excited-state energies and associated transition dipoles for 80,000 2T configurations harvested from the MD simulation of isolated 2T at 1000 K, and trained SchNet models for these properties. Following the same procedure described in the main text, we re-computed the absorption spectrum of 2T in dichloromethane including the first five excited states, and as shown in Fig. \ref{fig:t2}, a second peak shows up at around 5.8 eV, leading to an improved agreement between calculated and experimental spectra. Our calculation still underestimates the intensity of the high-energy peak, and possibly overestimates the peak position. More excited states may be still needed, but more importantly the errors associated with higher excited states may be even larger than that of the lowest-lying excited state. For 6T, our test calculation suggests that at least 30 excited states are needed to cover the spectral range up to 5.0eV. 

\renewcommand{\thetable}{S2}
\begin{table}[h!tbp]
\begin{tabular}{cccccc}
\hline
Solvent & 2T & 3T & 4T & 5T & 6T  \\
\hline
Chloroform\cite{Grebner1995}          & 4.11 & 3.50 & 3.18 & 2.98 & 2.87  \\
Dichloromethane\cite{Taliani}     & 4.09 & 3.50 & 3.15 & 2.96 & 2.83  \\
Dioxane\cite{Grebner1995,Lap1997}             & 4.05 & 3.49 & 3.16 & 2.99 & 2.85  \\
Hexane\cite{Sease1947}              & 4.12 & 3.54 & 3.22 & - & - \\
Benzene\cite{Sease1947}             & 4.07 & 3.49 & 3.17 & 2.97 & - \\
Benzene\cite{Zhao1988}             & 4.11 & 3.54 & 3.17 & 3.01 & 2.89 \\
\hline
\end{tabular}
\caption{Absorption maxima of the experimental absorption spectra of OTs in different solvents in the unit of eV.}
\label{tab:abs}
\end{table}

\renewcommand{\thefigure}{S1}
\begin{figure}[h!tbp]
  \centering
    \includegraphics[width=0.6\textwidth]{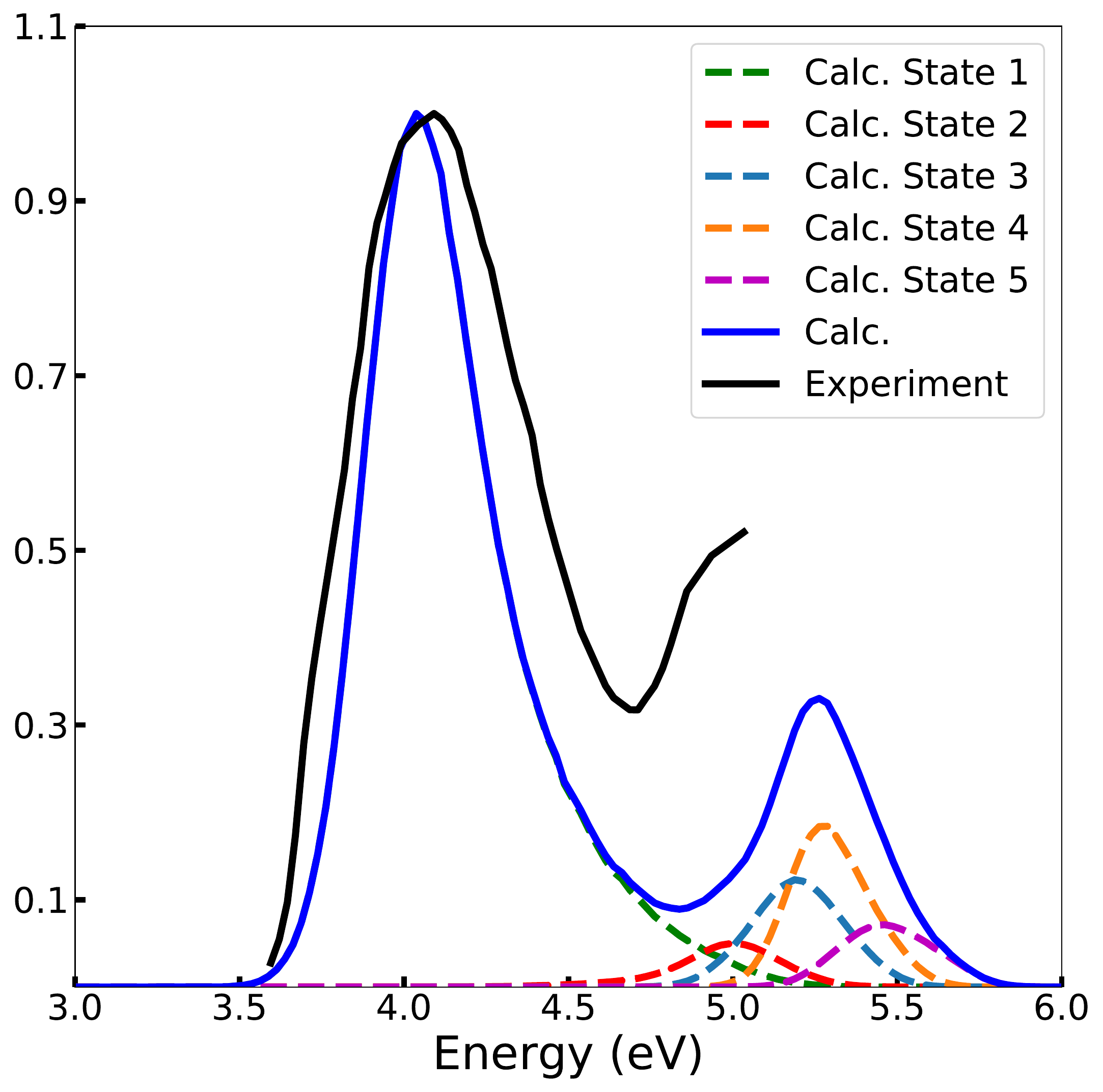}
     \caption{Calculated (blue solid line) and experimental\cite{Taliani} (black solid line) absorption spectra of 2T in dichloromethane. The contributions from the first five excited states to the calculated spectra are shown as colored dashed lines. All the calculated spectra are uniformly red-shifted by 0.3eV, and both the calculated and experimental spectra are scaled to have the same peak height.}
     \label{fig:t2}
\end{figure}

\renewcommand{\thefigure}{S2}
\begin{figure}[h!tbp]
  \centering
    \includegraphics[width=0.6\textwidth]{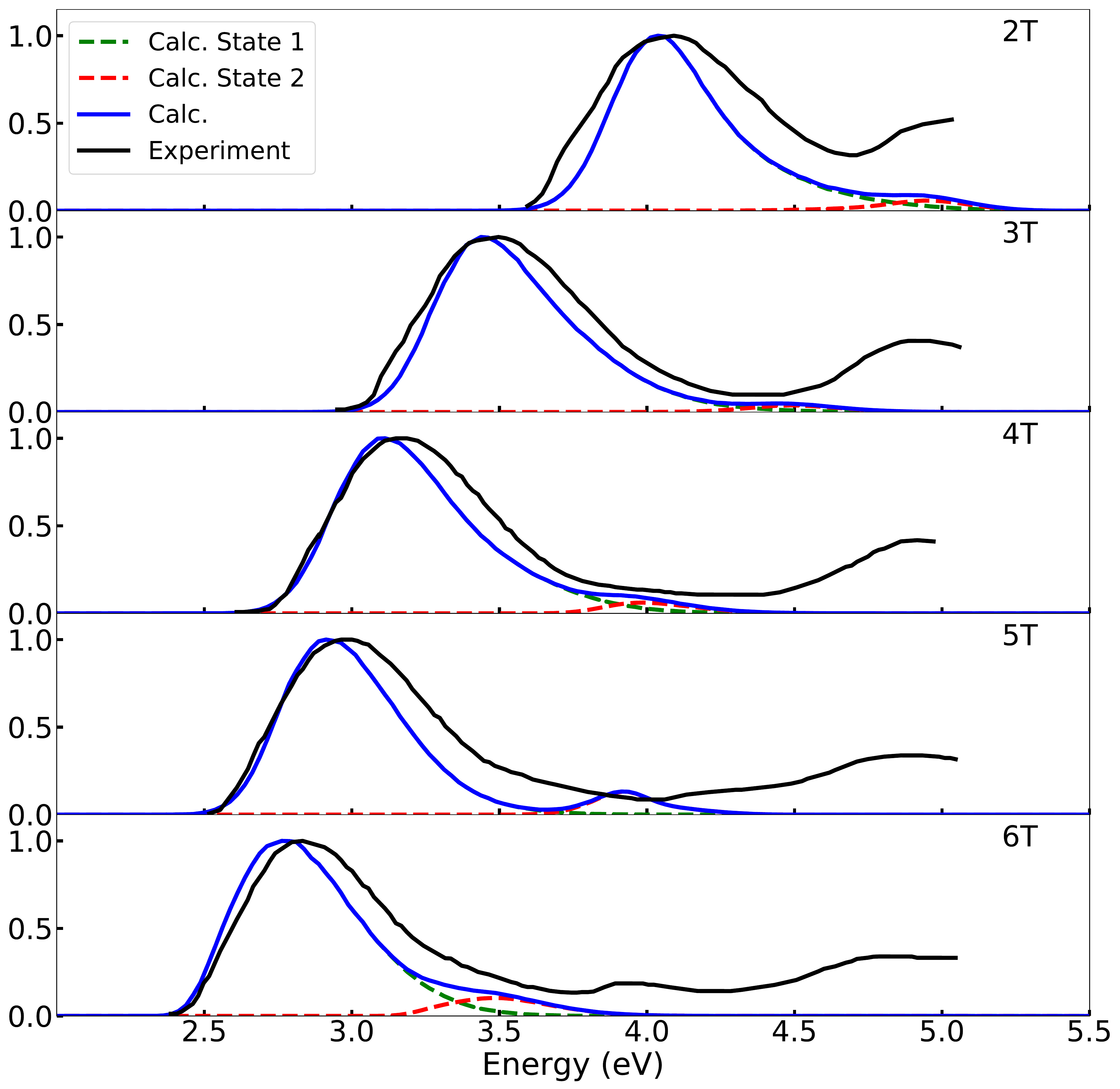}
     \caption{Calculated (blue solid lines) and experimental\cite{Taliani} (black solid lines) absorption spectra of OTs in dichloromethane. The contributions from the first two excited states to the calculated spectra are shown as green and red dashed lines, respectively. All the calculated spectra are generated from SchNet models trained on 5000 configurations, and are uniformly red-shifted by 0.3eV. Both the calculated and experimental spectra are scaled to have the same peak height.}
     \label{fig:5k}
\end{figure}

In the main text, the SchNet models for excited-state energies and associated transition dipoles are trained on 80,000 configurations for each OT. We repeated the training process but with only 5,000 randomly chosen configurations, and the resulting absorption spectra are displayed in Fig. \ref{fig:5k}. The agreement between calculated and experimental spectra is almost as good as that based on 80,000 training configurations (see Fig. 6 in the main text) except two minor differences: the spectra are slightly red-shifted; and the spectra of 5T and 6T are slightly narrower. This is consistent with the dependence of MAEs on training data size shown in Fig. 2 of the main text, demonstrating the effectiveness of SchNet in predicting excited-state properties of OSCs even with relatively small datasets.

\clearpage
\bibliographystyle{naturemag_noURL}
\bibliography{MyCollection}